\documentclass[]{aastex}

\usepackage{emulateapj5}
\usepackage{onecolfloat}
\usepackage{graphicx}
\usepackage{fancyheadings}
\usepackage{ulem}
\usepackage{rotating}
\usepackage{lscape}

\def\NH{$N$(HI)}

\def\etal{et al.}

\def\lya{Ly$\alpha$ }
\def\smpykpc{M$_{\odot}{\rm \ yr^{-1} \ kpc^{-2}}$}

\def\rvec{${\bf r}$}

\def\kms{km~s$^{-1}$ }

\def\micron{$\mu$m}

\def\gamdnr{${\Gamma_{d}}$}

\def\taunu{$\tau_{\nu}$}

\def\kapnr{${\kappa}$}
\def\cm2{\, \rm cm^{-2}}

\def\perd{\;\;\; .}
\def\cmma{\;\;\; ,}

\def\rhodot{$\dot{\rho_{*}}$}

\def\ps{$\dot{\psi_{*}}$}

\def\psav{$<$$\dot{\psi_{*}}$$>$}
\def\psavz{${<{{\dot{\psi_{*}}}}(z)>}$}
\def\ciis{C II$^{*}$}
\def\nh{$N$(H I)}
\def\lclos{$l_{c}$}
\def\lcav{$<l_{c}>$}
\def\lcr{$l_{cr}({\rm {\bf r}})$}
\def\lcrnr{$l_{cr}$}
\def\jnu{$J_{\nu}$}

\begin{document}

\twocolumn[%
\submitted{submitted to the Astrophysical Journal Nov.5,2002}

\title{CII$^{*}$ ABSORPTION IN DAMPED \lya\ SYSTEMS: (I)
STAR FORMATION RATES IN A TWO-PHASE MEDIUM}

\author{ ARTHUR M. WOLFE,\altaffilmark{1}\\ 
Department of Physics and Center for Astrophysics and Space Sciences; \\
University of California, San
Diego; \\
C--0424; La Jolla, CA 92093\\
{\bf awolfe@ucsd.edu}}
 
\author{} 

\author{JASON X. PROCHASKA,\altaffilmark{1}\\ 
UCO-Lick Observatory; \\
University of California, Santa Cruz\\
Santa Cruz, CA; 95464\\
{\bf xavier@ucolick.org}}

\author{and}

\author{ ERIC GAWISER\altaffilmark{1,2}\\ 
Department of Physics and Center for Astrophysics and Space Sciences; \\
University of California, San
Diego; \\
C--0424; La Jolla, CA 92093\\
{\bf egawiser@ucsd.edu}}

\begin{abstract}

We describe a technique that for the first time measures
star formation rates (SFRs) in damped Lyman alpha systems
(DLAs) directly. We assume that massive stars form in DLAs,
and that the FUV radiation they emit  
heats the gas by the grain photoelectric mechanism. 
We infer
the heating rate by equating it to the cooling rate measured by the
strength of CII$^{*}$ 1335.7 absorption. 
Since the heating rate is proportional to the product of the
dust-to-gas ratio, the grain photoelectric heating efficiency, 
and the  SFR per unit area,  
we can deduce the SFR per unit area
for DLAs in which the cooling rate and the dust-to-gas ratio have been
measured. We consider models in which the dust consists of
carbonaceous grains and silicate grains.
We present two phase models in which the 
cold neutral medium (CNM) and
warm neutral medium (WNM) are in pressure equilibrium. In the CNM model
the line of sight passes through CNM and WNM gas, while
in the WNM model the line of sight passes only through 
WNM gas. Since the grain photoelectric heating efficiency
is at least an order of 
magnitude higher in the CNM than in the WNM, most of
the CII$^{*}$ absorption arises in the CNM in the CNM model.
By contrast, in the WNM model all of the CII$^{*}$ absorption arises 
in the WNM. We use the measured CII$^{*}$ absorption
lines to derive  the SFR per unit area
for a sample of $\approx$ 30  DLAs in which the dust-to-gas ratio 
has been inferred from element depletion patterns.
We show that the inferred SFR per unit area corresponds to an average over
the star forming volume of the DLA rather than to local
star formation along
the line of sight. We obtain the average SFR
per unit area and show that it equals 
10$^{-2.2}$ M$_{\odot}$yr$^{-1}$kpc$^{-2}$
for the CNM solution and 10$^{-1.3}$ 
M$_{\odot}$yr$^{-1}$kpc$^{-2}$
for the WNM solution. Interestingly, the SFR
per unit area in the 
CNM solution is similar
to that measured in the Milky Way ISM.

 \end{abstract}


]
\altaffiltext{1}{Visiting Astronomer, W.M. Keck Telescope.
The Keck Observatory is a joint facility of the University
of California and the California Institute of Technology.}

\altaffiltext{2}{Current address: Yale Astronomy Department,
P. O. Box 208101, New Haven, CT, 06520}

\pagestyle{fancyplain}
\lhead[\fancyplain{}{\thepage}]{\fancyplain{}{Wolfe, Prochaska, \& Gawiser}}
\rhead[\fancyplain{}{STAR FORMATION IN DAMPED {\lya} SYSTEMS}]{\fancyplain{}{\thepage}}
\setlength{\headrulewidth=0pt}
\cfoot{}

\section{INTRODUCTION}

Star formation is a key ingredient in the formation and evolution of galaxies.
The idea  
that the Hubble sequence is 
actually a sequence in present-day star formation rates (SFRs)
and past star formation histories (Roberts 1963) is supported by
results of population synthesis models
(Tinsley 1980) and by the use of  
precise diagnostics of SFRs such as emission-line
fluxes (Kennicutt 1983), and UV continuum luminosities. 
According to this interpretation,
late-type Sc galaxies are characterized by low SFRs that are
independent of time, while in early-type Sa galaxies an initially high
SFR decreases steeply
with time (Kennicutt 1998). 
Star formation
may also influence galaxy evolution  through feedback. That is,
shocks generated by 
supernova explosions heat gas in protogalaxies, thereby delaying
cooling and collapse to rotating disks. This process has been invoked
in hierarchical cosmologies in order to
prevent the collapse of too many baryons into low-mass dark matter halos,
and to explain the angular momentum distribution of current galaxy disks
(Efstathiou 2000).

A {\em direct} measurement of star formation
in high redshift galaxies would provide an independent test of these ideas. 
Madau {\etal} (1996) took a crucial first step in this direction when they reconstructed
star formation histories
by measuring the comoving luminosity density of star-forming galaxies
as a function of redshift. While the original ``Madau diagrams''
exhibited a peak in star formation at $z$ ${\approx}$ 1$-$2, more recent
analyses, which are based on larger samples of galaxies and correct for extinction
of emitted starlight by dust, show no evidence for such a peak. Rather,
the SFR per unit comoving volume, {\rhodot}, increases by
a factor of $\sim$10 in the redshift interval $z$ =[0,1] and either remains constant
out to the highest redshifts confirmed so far $z$ $\approx$ 6 (Steidel {\etal} 1999) or
keeps increasing to even higher redshifts (Lanzetta {\etal} 2002).
However, the galaxies from which these results
are derived are {\em unlikely} to be the progenitors of the bulk of the 
current galaxy population.  
Whereas the star formation rate per unit area  for the Milky Way Galaxy is given by
{\ps} $\sim$ 4$\times$10$^{-3}$ {\smpykpc} (Kennicutt 1998), the comoving SFR
at $z$ $\sim$ 3 is inferred from
the Lyman Break galaxies, a highly luminous population of star forming objects in
which
{\ps} $\ge$ 1 {\smpykpc} (Pettini {\etal} 2001). The Galaxy was unlikely to
be this bright in the past, as
stellar population studies predict that at $z$ $\sim$ 3,
{\ps} $\sim$ 2$\times$10$^{-2}$ {\smpykpc} for
Sb galaxies such as the Milky
Way. 
In fact only ellipticals are predicted to have 
{\ps} $\sim$ 1 {\smpykpc} at $z$ $\sim$ 3 (see
Genzel {\etal} 2001). This is consistent with other independent 
lines of evidence
such as strong clustering (Adelberger {\etal} 1998), which suggests
the Lyman Break galaxies evolve into massive
ellipticals in rich clusters. 
As a result, the published Madau diagrams need not 
constrain the star formation history of normal 
galaxies or their progenitors.

This is the first of two papers in which
we present a new technique for measuring SFRs in
DLAs. 
The idea is to infer the SFR from the rate at which neutral gas in DLAs is heated.
We determine the
heating rate by equating it to the cooling rate; i.e., we assume
steady-state conditions (see $\S$ 4). We estimate the cooling rate 
from  [C II] 158 $\mu$m emission, the dominant coolant
for the Milky Way ISM (Wright {\etal} 1991). Specifically, we measure
the cooling rate per H atom
from the strength of {\ciis} 1335.7 absorption. As we shall show,
it is plausible to assume
the heating rate is proportional to the mean intensity of far UV (FUV)
radiation, which is proportional to {\ps} for a plane parallel layer.
The goal of this paper is to determine
the mean star formation rate per unit physical area, 
 {\psavz}, 
for a given redshift bin.
In the second paper (Wolfe {\etal} 2003; hereafter Paper II)
we combine {\psavz} with the incidence of DLAs per unit redshift
to obtain {\rhodot}($z$) for our DLA sample. The advantage of our method
is this: Because DLAs are not drawn
from a flux limited sample of galaxies,
we are able to 
derive
{\ps} values far below those
determined from radiation emitted by rapidly star forming objects
such as Lyman Break galaxies. 
Rather, we will show that our technique is
sensitive to {\ps} as low as 1$\times$10$^{-4}$ {\smpykpc};
i.e., to SFRs per unit area below the flux thresholds of detectors
on 10 m class telescopes.

A further advantage of our technique is that
it probes the physical state of neutral gas at high redshifts.
Specifically, from our determination of the heating rate 
we compute
the thermal equilibrium of the neutral gas.  By analogy to models for the ISM
(here and throughout the paper ISM refers to the interstellar medium
of the Milky Way Galaxy)
our calculations predict
a two-phase medium comprising 
a cold ($T$ $\sim$ 80 K) neutral medium (CNM) and
a warm ($T$ $\sim$ 8000 K)
neutral medium (WNM) in pressure equilibrium with
each other.
We consider two models: one in which the line of 
sight to the QSO encounters gas in both CNM and WNM phases
and another in which it encounters only the
WNM phase. The WNM model is considered because
of recent arguments that DLAs consist only
of WNM gas
(Norman \& Spaans 1997; Liszt 2002;
and Kanekar \& Chengalur 2001).
We find that the technique is insensitive
to the masses and  sizes of individual DLAs. This has the advantage that
the results are not critically dependent on model assumptions such as
the mass or length scale of the dark matter mass distribution
(e.g. Prochaska \& Wolfe 1997; 
Haehnelt {\etal} 1998).

The paper is organized as follows. In $\S$ 2 we describe
the basic strategy for inferring the SFR per unit area from  
measurements of {\ciis} 1335.7 absorption lines. 
We present data for 33 DLAs obtained mainly with the
HIRES echelle spectrograph (Vogt {\etal} 1994) on the Keck I 10 m 
telescope in $\S$ 3. $\S$ 4 presents
two-phase models for the neutral gas in DLAs.
We explain how radiative excitations cause the [C II] 158 {\micron}
emission rate to exceed the 158 {\micron} cooling rate.
In $\S$ 5 solutions to the transfer equation for
FUV radiation
are given. We use these solutions to predict heating rates as functions of {\ps} and
dust-to-gas ratio. For each DLA we measure {\ps} by combining
measurements of heating rate and dust-to-gas 
ratio with the solutions. 
We then determine  $<${\ps}$>$
for two redshift bins drawn from the full 
sample of 33 DLAs. The significance of these
measurements is they refer to objects representative of the protogalactic
mass distribution; i.e., objects likely to evolve into normal galaxies.
In $\S$ 6 we test the assumptions of our dust models
for self consistency.
A summary and concluding remarks are presented in $\S$ 7.

\section{THE IDEA}

Our technique is based on the idea that massive stars form
out of gas in DLAs. Evidence for this stems from 
the physical resemblance between  DLAs and the neutral gas of the ISM,
the presence
of heavy elements in DLAs, and the fact that
DLAs in the redshift interval $z$ = [2,3] contain sufficient baryons 
in the form of neutral gas to
account for all the visible stars in current spiral galaxies
(Storrie-Lombardi \& Wolfe 2000). Although
stars likely form out of molecular rather than atomic gas,
and molecules are rarely detected in DLAs (Lu {\etal} 1997;
Petitjean {\etal} 2000),  the presence of heavy elements argues
for the formation of  
massive stars in DLAs.
Such stars emit FUV radiation ($h{\nu} = 6 - 13.6$ eV) that illuminates dust grains
known to be present in the gas (Pei \& Fall 1995; Prochaska \&
Wolfe 2002; hereafter PW02). By analogy with the
Milky Way ISM,
a small fraction of the incident photon energy is transferred
to photo-ejected electrons that heat the gas via Coulomb interactions with
ambient electrons (e.g. Bakes \& Tielens 1994; Wolfire {\etal} 1995 [hereafter W95]). In this
case the heating rate per H atom at displacement
vector {\rvec} is given by
\begin{equation}
 {\Gamma}_{d}({\bf r}) = 10^{-24}{\kappa({\bf r})}{\epsilon}{G_{0}}({\bf r}) \ {\rm ergs \ s^{-1} \ H^{-1}}
\perd
\label{eq:Gamddef}
\end{equation}

\noindent In the last equation
${\kappa}$({\rvec})${\equiv}$$k$({\rvec})$_{DLA}$/$k_{MW}$, where $k$({\rvec})$_{DLA}$
is the dust-to-gas ratio in the DLA at {\rvec}, and $k_{MW}$ is the
dust-to-gas ratio assumed 
for the present epoch ISM of the Galaxy
(Bakes \& Tielens 1994; see discussion in $\S$ 4.1). 
The incident FUV radiation field {$G_{0}$} equals 4${\pi}J$, 
where $J$ is the mean intensity integrated between 6 and 13.6 eV,
and 
is in units of Habing's (1968) estimate of the local interstellar
value (=1.6$\times$10$^{-3}$ ergs cm$^{-2}$ s$^{-1}$).
The quantity 
$\epsilon$ is the fraction of FUV radiation absorbed by grains
and converted to gas heating (i.e., the heating efficiency); $\epsilon$ 
is also a function of $G_{0}T^{1/2}/n_{e}$ (Bakes \& Tielens 1994).
For a plane parallel layer, $J$ is proportional to  the source luminosity
density projected perpendicular to the plane; i.e., the
source luminosity per unit area,
which in the case of FUV radiation is proportional to {\ps} ($\S$ 5.1). As a result, 
we can deduce {\ps} provided we know {\kapnr}, {\gamdnr}, and 
{$\epsilon$}. Because
$\epsilon$ is well determined for a wide range 
of physical conditions
(Bakes \& Tielens 1994; Weingartner \& Draine 2001a),
this reduces to measuring the heating rate
and {$\kappa$}. Both of these are obtainable
from HIRES spectroscopy since the heating rate can be
inferred from
{\ciis} absorption and {\kapnr} is determined 
from the abundance patterns 
and metallicity of the gas
(see $\S$ 4.1). Note, to derive equation (1) we assume {$\epsilon$}
in high-$z$ DLAs is the same as in the ISM. In other words we
assume quantities determining $\epsilon$, such
as the photoelectric cross-section, photo-electric ionization yield,
kinetic energy partition function, and grain size distribution 
(Bakes \& Tielens 1994) are
the same in DLAs and the ISM.

We determine the heating rate by equating it to the cooling rate; i.e,
we assume steady-state conditions.
This is reasonable since the cooling times  are $\approx$ 10$^5$$-$10$^{6}$  yrs
which are short compared to the dynamical time scales for most model
protogalaxies (see $\S$ 4.3). As a result we let 
\begin{equation}
 {\Gamma} = n{\Lambda}
\cmma
\label{eq:Gameqnlam}
\end{equation}

\noindent where the total heating rate $\Gamma$
includes other sources of heat in addition to {\gamdnr}, and
$n$ and $\Lambda$ are the gas density and cooling function.
In the ISM, cooling is dominated by [C II] 158 {\micron}
emission, i.e., $\Lambda_{{\rm CII}}$,  with a luminosity \\ 
$L$([C II]) = 5$\times$10$^{7}$$L_{\odot}$
(Wright {\etal} 1991). The 158 {\micron} line results 
from transitions between the  
$^{2}P_{3/2}$
and $^{2}P_{1/2}$ fine-structure states in the ground 
2$s^{2}$2$p$  term of C$^{+}$. 
Most of the emission from the Galaxy and other nearby spirals arises
in the diffuse CNM gas rather than from star-forming regions in spiral arms,
or photo-dissociation regions on the surfaces
of molecular clouds (e.g. Madden {\etal} 1993). The last point is
especially relevant for damped {\lya} systems where molecules are
rarely detected (Lu {\etal}
1997; Petitjean {\etal} 2000).

Pottasch {\etal} (1979) used the following expression 
to estimate the [C II] 158 {\micron} emission per H atom from 
gas detected in absorption against background sources:
\begin{equation}
l_{c} = N({\rm C II}^{*})h{\nu}_{ul}A_{ul}/N({\rm H I}) \ {\rm ergs \ s^{-1} \ H^{-1}}
\cmma
\label{eq:lceqNCIIovNH}
\end{equation}

\noindent where  $N$({\ciis}) is the column density of C$^{+}$  ions in the $^{2}P_{3/2}$
state, {\nh} is the H I column density, and $A_{ul}$  
and $h{\nu}_{ul}$ 
are the coefficient for spontaneous photon decay
and energy of the $^{2}P_{3/2}$ $\rightarrow$ $^{2}P_{1/2}$ transition.
In fact, {\lclos} is just the density$-$weighted average along the line of sight of the
more fundamental quantity, {\lcr}, the rate of spontaneous emission
of energy per H atom at a given displacement
vector ${\bf {r}}$. That is
\begin{equation}
l_{c} = {{\int n_{{\rm HI}}(s)l_{cr}({\rm {\bf r}(s)})ds} \over {\int n_{{\rm HI}}(s)ds}}
\cmma
\label{eq:lcavlcr}
\end{equation}

\noindent

\noindent where
\begin{equation}
l_{cr}({\rm {\bf r}}) = {{n_{{\rm CII}^{*}}({\rm {\bf r}})A_{ul}h{\nu}_{ul}} \over { n_{{\rm HI}}({\rm {\bf r
}})}}
\label{eq:lcrdef}
\cmma
\end{equation}

\noindent $n_{{\rm CII}^{*}}$ and $n_{{\rm H I}}$ are the volume densities of H I and {\ciis},
and $ds$ is the differential path length along the line of sight.
Notice that {\lcrnr}=4{$\pi$}$j/n_{{\rm H I}}$ where $j$ is the volume
emissivity appearing in the radiative transfer equation.
We can measure {\lclos} since $N$({\ciis}) and {\nh} are measurable:
$N$({\ciis}) from {\ciis} 1335.7 absorption and {\nh} from damped
{\lya} 1215.7 absorption. {\em Note also that {$\Gamma$} = {\lcrnr} when cooling is dominated by
{\rm [C II]} 158 {\micron} emission}.

\section{THE DATA}

\begin{table*}[ht] \footnotesize 
\begin{center}
\caption{{\sc DLA PROPERTY}}
\begin{tabular}{lcccccc}
\tableline
\tableline
\cline{3-7}
    &           &log$_{10}${\nh} & log$_{10}N$({\ciis})&[Fe/H] & [Si/H] &  log$_{10}${\lclos}  \\
QSO$^{a}$ & $z_{abs}$ & (cm$^{-2})$&         (cm$^{-2}$)&      &        & (ergs s$^{-1}$ H$^{-1}$)  \\
(1)       & (2)       & (3)        &         (4)        &  (5) &  (6)   & (7)                       \\
\tableline
Q0019$-$15 & 3.439 & 20.92$\pm$0.10 & 13.84$\pm$0.02 & $-$1.587$\pm$0.108 &$-$1.058$\pm$0.113 &$-$26.61$\pm$0.10 \cr
Q0100$+$13${^b}$   & 2.309 & 21.37$\pm$0.05 & 13.59$\pm$0.05 & $-$ 1.899$\pm$0.090 &$-$1.460$\pm$0.081& $-$27.33$\pm$0.07 \cr
Q0149$+$33   & 2.141 & 20.50$\pm$0.10 & $<$ 12.78 & $-$ 1.770$\pm$0.102 &$-$1.489$\pm$0.110& $< \ -$27.24 \cr
Q0201$+$11$^{c,h}$ & 3.387 & 21.26$\pm$0.10 &  14.12$\pm$0.10&$-$ 1.410$\pm$0.113 &$-$1.250$\pm$0.150& $-$26.67$\pm$0.10 \cr
Q0255$+$00   & 3.915 & 21.30$\pm$0.05 & 13.44$\pm$0.04  & $-$2.050$\pm$0.101 &$-$1.779$\pm$0.052& $-$27.38$\pm$0.07 \cr
Q0307$-$49$^{d}$   & 4.466 & 20.67$\pm$0.09 & $<$ 13.59&$-$1.960$\pm$0.220&$-$1.550$\pm$0.12&$< \ -$26.60 \cr
Q0336$-$01   & 3.062 & 21.20$\pm$0.10 &14.04$\pm$0.03& $-$1.795$\pm$0.105&$-$1.406$\pm$0.100&$-$26.68$\pm$0.10 \cr
Q0347$-$38   & 3.025 & 20.63$\pm$0.10 & 13.47$\pm$0.03&$-$1.623$\pm$0.080&$-$1.170$\pm$0.026& $-$26.68$\pm$0.11 \cr
Q0458$-$02   & 2.039 & 21.65$\pm$0.09  &$>$ 14.80&$-$1.767$\pm$0.102 &$-$1.186$\pm$0.092& $> \ -$26.38 \cr
Q0741$+$47   & 3.017&20.48$\pm$0.10&$<$ 12.55&$-$1.928$\pm$0.100&$-$1.686$\pm$0.100&  $< \ -$27.45\cr
Q0836$+$11   & 2.465 & 20.58$\pm$0.10 &$<$ 13.12&$-$1.403$\pm$0.101 &$-$1.154$\pm$0.110&  $< \ -$26.98\cr
Q0951$-$04$^{g}$   & 4.203 & 20.40$\pm$0.10 & 13.37$\pm$0.08&$< \ -$2.591&$-$2.618$\pm$0.104&   $-$26.55$\pm$0.13\cr
Q0952$-$01$^{g}$   & 4.024 & 20.55$\pm$0.10 & 13.55$\pm$0.02 &$-$1.863$\pm$0.126 &---& $-$26.52$\pm$0.10\cr
Q1104$-$18   & 1.661 & 20.80$\pm$0.10& 13.44$\pm$0.05 &$-$1.476$\pm$0.102 &$-$1.040$\pm$0.100&$-$26.93$\pm$0.11 \cr
Q1108$-$07   & 3.608 & 20.50$\pm$0.10 & $<$ 12.34&$-$2.116$\pm$0.101&$-$1.798$\pm$0.100&$< \ -$27.68\cr
Q1202$-$07$^{b}$   & 4.383 & 20.60$\pm$0.14 & $<$ 13.06&$-$2.19$\pm$0.188 &$-$1.809$\pm$0.141& $< \ -$27.06\cr
Q1215$+$33   & 1.999 & 20.95$\pm$0.067 & $<$ 13.17&$-$1.702$\pm$0.085&$-$1.481$\pm$0.072&$< \ -$27.30\cr
Q1223$+$17   &2.466  & 21.50$\pm$0.10 & $<$ 14.01&$-$1.843$\pm$0.102 &$-$1.593$\pm$0.100& $< \ -$27.02\cr
Q1232$+$08$^{e}$   &2.337  & 20.90$\pm$0.10&14.00$\pm$0.10&$-$1.720$\pm$0.13 &$-$1.220$\pm$0.15&$-$26.40$\pm$0.14\cr
Q1331$+$17$^{f}$   &1.776  & 21.18$\pm$0.04&14.05$\pm$0.05&$-$2.058$\pm$0.41 &$-$1.452$\pm$0.041& $-$26.65$\pm$0.07\cr
Q1346$-$03   &3.736  & 20.72$\pm$0.10& 12.55$\pm$0.11 &$-$2.634$\pm$0.102&$-$2.332$\pm$0.100&$-$27.69$\pm$0.15\cr
Q1425$+$60$^{g}$   &2.827  & 20.30$\pm$0.04&$<$ 13.33&$-$1.329$\pm$0.040&$>$ $-$1.034&$< \ -$26.50\cr
Q1443$+$27$^{g}$   &4.224  & 20.80$\pm$0.10 & $>$ 14.71&$-$1.096$\pm$0.115 &$> \ -$0.926&$> \ -$25.61\cr
Q1759$+$75$^{g}$   &2.625  & 20.76$\pm$0.10 &12.80$\pm$0.05 &$-$1.184$\pm$0.008&$-$0.786$\pm$0.011&$-$27.52$\pm$0.11 \cr
Q1946$+$76$^{b}$  &2.844 &20.27$\pm$0.06&$<$ 12.46&$-$2.528$\pm$0.061 &$-$2.226$\pm$0.060& $< \ -$27.33 \cr
Q2206$-$19   &2.076  & 20.43$\pm$0.06 & $<$ 13.16&$-$2.606$\pm$0.062&$-$2.309$\pm$0.069&$< \ -$26.80 \cr
Q2231$-$00   &2.066  & 20.56$\pm$0.10 & 13.71$\pm$0.04 &$-$1.402$\pm$0.119&$-$0.875$\pm$0.102& $-$26.38$\pm$0.11 \cr
Q2237$-$06$^{b}$   &4.080&20.52$\pm$0.11&$<$ 12.53& $-$2.140$\pm$0.167&$-$1.870$\pm$0.112&$< \ -$27.51 \cr
Q2343$+$12$^{b,g}$   &2.431  & 20.34$\pm$0.10 & 12.77$\pm$0.05  &$-$1.199$\pm$0.100 &$-$0.540$\pm$0.101&$-$27.09$\pm$0.11 \cr
Q2344$+$12$^{b,h}$   &2.538  & 20.36$\pm$0.10 &$<$ 12.95&$-$1.830$\pm$0.105 &$-$1.741$\pm$0.101&$< \ -$26.93 \cr
Q2348$-$14   &2.279  & 20.56$\pm$0.08&$<$ 13.21&$-$2.238$\pm$0.0770&$-$1.917$\pm$0.0776&$< \ -$26.88 \cr
Q2359$-$02A  &2.095  & 20.70$\pm$0.10&  13.70$\pm$0.06 &$-$1.655$\pm$0.103&$-$0.778$\pm$0.102& $-$26.52$\pm$0.12 \cr
Q2359$-$02B  &2.154  & 20.30$\pm$0.10 & $<$ 14.48 & $-$1.877$\pm$0.105 &$-$1.583$\pm$0.101&$< \ -$25.34 \cr

\tableline
\end{tabular}
\end{center}


\tablenotetext{a}{Data from UCSD Data base unless otherwise noted}
\tablenotetext{b}{Data from Lu {\etal} (1997)}
\tablenotetext{c}{Data from 
Ellison {\etal} (2001)}
\tablenotetext{d}{Data from Dessauges-Zavadsky {\etal} (2001).}

\tablenotetext{e}{Data from Srianand {\etal} (2000).}

\tablenotetext{f}{$N$({\ciis}) from Songaila \& Cowie  (2001).}

\tablenotetext{g}{Excluded from minimal and maximal depletion models (see $\S$ 5.2.2)}

\tablenotetext{h}{Excluded only from minimal depletion model (see $\S$ 5.2.2)}

\end{table*}

We have determined {\lclos} for 33 DLAs. The results
were obtained
by measuring $N$({\ciis}) and 
{\nh} from accurate velocity profiles. 
In Table 1 we show {\lclos}  and other properties
to be used in subsequent analyses.
Column 1
gives the coordinate name of the background QSO, column 2 the absorption redshift
of the DLA, column 3 the H I column density, column 4 the {\ciis} column
density, column 5 the iron abundance relative to solar
(where [Fe/H]$\equiv$log$_{10}$(Fe/H)
$-$log$_{10}$(Fe/H)$_{\odot}$), column 6 gives the silicon abundance, [Si/H],  
and {\lclos} is given in 
column 7. In cases  where Fe absorption lines were not measured,
we  substituted proxy elements such as Ni, Cr, and Al. In cases
where Si absorption was not measured, we used S or Zn as proxies
(see PW02 for a full description of these procedures).
The data for 30 of the entries were obtained
with 
HIRES: 23 of these by our group (Prochaska {\etal} 2001), 6 by
Lu {\etal} (1997), and one by Songaila \& Cowie (2001).
Data for the remaining 3 entries were acquired
with the UVES spectrograph on the VLT 8 m telescope (Ellison {\etal} 2001;
Srianand {\etal} 2000; Dessauges-Zavadsky {\etal} 2001).

Figure 1  shows six
examples of {\ciis} velocity profiles used to derive $N$({\ciis}) 
in Table 1 along with corresponding low-ion resonance
profiles. While the velocity structures of the two profiles exhibit
overall similarity, statistically significant differences exist. These
are evident in (1) the DLA
toward Q0347$-$38 (panel c) where
two strong velocity components are detected at $v$=$-$8 and $+$12 {\kms}
in Fe II 1608, while only the $v$=$+$12{\kms} component is detected
in the {\ciis} 1335.7 profile even though there is sufficient 
signal-to-noise to detect the $v$=$-$8{\kms} component,
and (2) the DLA toward Q2231$-$00 (panel f) where absorption in {\ciis} 1335.7 
between $-$50 {\kms} and $-$20 {\kms} is not detected in
Si II 1808. In Paper II we show these differences to be evidence
for a multi-phase gas in which {\lcr} varies along the line of sight.

In Figure 2 we plot {\lclos} versus  {\nh} for the DLAs.  
There  are
16 positive detections (red data points), 2 lower limits 
(95$\%$ confidence intervals plotted as blue points), and 15  upper
limits (95$\%$ confidence intervals plotted as green data points). 
One purpose of this plot is to
illustrate possible systematic effects such as correlations between {\lclos} and
 \\ {\nh}.
No such correlation 
is evident in the data. However, 
there is a tendency for the upper limits on {\lclos} to
occur at low H I column densities:  12 of the 15 upper limits
on $N$({\ciis}) occur
at log$_{10}${\nh} $\le$ 20.6 cm$^{-2}$. This suggests  that at least
some of the null detections with large upper limits 
result from gas with  
{\ciis} column densities sufficiently low
that 1335.7 is undetectable 
rather than from low
values of {\lclos}.
Other upper limits are caused by blending
between {\ciis} 1335.7 and {\lya} forest lines (as for the $z$ = 2.154 DLA
toward Q2359$-$02B). However, we cannot exclude the  possibility
that the remaining upper limits arise from
{\lclos} values substantially
below the positive detections.
The lower limits correspond to cases in which
{\ciis} 1335.7 is saturated.

\begin{figure}[ht]
\includegraphics[height=3.8in, width=3.7in]{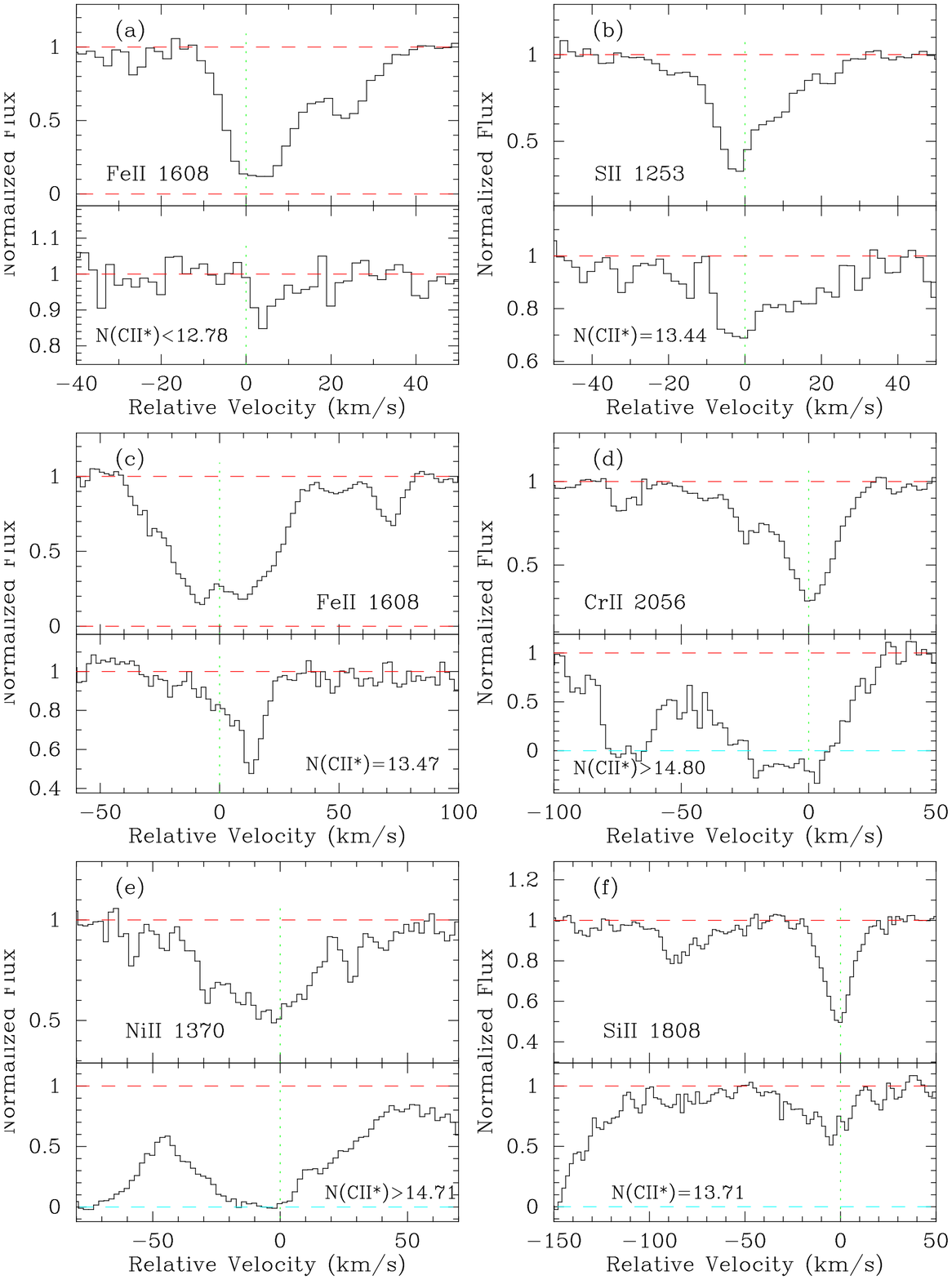}
\caption{Velocity profiles comparing {\ciis} and selected resonance lines
in 6 DLAs. In each case the resonance transition and {\ciis} column
density, $N$({\ciis}), are specified.}
\label{ciistarprofile}
\end{figure}

\begin{figure}[ht]
\includegraphics[height=3.8in, width=3.7in]{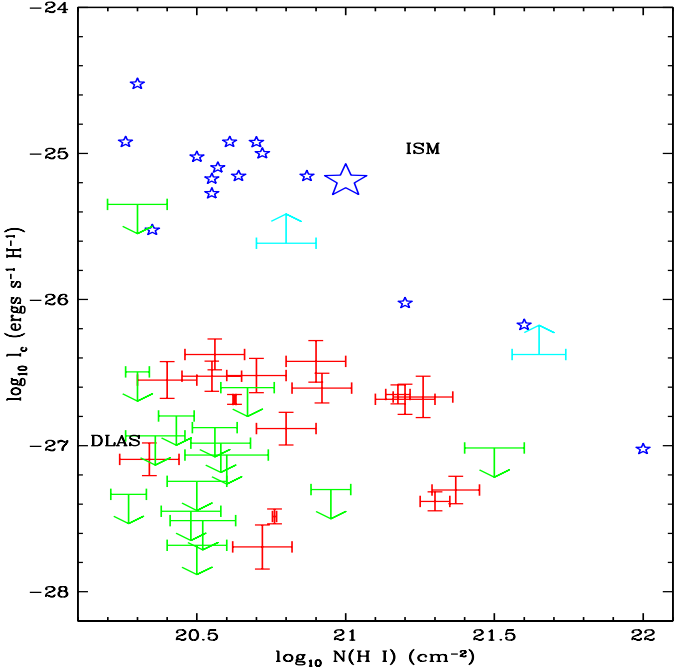}
\caption{Inferred [C II] spontaneous emission rate per H atom, $l_{c}$,
versus H I column density. Red data points are positive detections with
error bars for 16 DLAs. Green data points show 2-$\sigma$ upper limits
on $l_{c}$ for 15 DLAs, and blue data points depict 2-$\sigma$
lower limits on $l_{c}$ for 2 DLAs. Small blue stars depict detections
toward H I clouds in ISM. Large blue star is spontaneous emission rate
per H atom averaged over 
disk of 
Galaxy}
\label{lcvsNHI}
\end{figure}

The second purpose of this plot is to
compare {\ciis} emission rates in DLAs and in the ISM.
Thus we plot {\lclos}, {\nh} pairs  derived for representative sightlines
through the ISM which are shown as small blue stars 
(Pottasch {\etal} 1979; Gry {\etal} 1992). While Pottasch {\etal} (1979)
did not report measurement errors, Gry {\etal}
(1992) report 1-$\sigma$ errors
corresponding to $\approx$ 0.3 dex
in log$_{10}$({\lclos}).
The large star  was derived by dividing the total 158 {\micron}
luminosity of the Galaxy by the H I mass of the disk 
(Hollenbach \& Tielens 1999); the result
corresponds to the density-weighted average of {\lcr} integrated over the
disk of the Galaxy which we refer to as  $<$$(l_{c})$$_{ISM}$$>$.
Comparison between the DLA and ISM data demonstrates that
{\lclos} averaged over the DLA sample, i.e., {\lcav}, is about 1/30 times
$<$$(l_{c})$$_{ISM}$$>$. Because
158 {\micron} emission from DLAs has not been detected, 
the analogous quantity, which is 
the density-weighted average of {\lcr} over
the  entire H I mass distribution of the DLAs, is unknown. Nevertheless
the data covering 33 sightlines through DLAs  strongly suggest the
{\ciis} cooling rate per H atom to be much lower in DLAs than
in the ISM. The heating
rates are therefore correspondingly lower.

The ratio of the two heating rates is simply 
explained if the DLA gas is heated by the same mechanism
that heats the ISM; i.e., photoejection of electrons
from dust grains. 
If the mean intensities of FUV radiation
are the same and the photoelectric efficiencies
are the same, equation 1 shows that the ratio of heating rates equals the
ratio of dust-to-gas ratios, $k_{DLA}/k_{MW}$.
Pettini {\etal} (1994) set $k_{MW}$ equal to the 
mean value in the Galaxy and
estimate that
$k_{DLA}/k_{MW}$ $\approx$ 1/30 to 1/20 which
approximates the DLA metallicity relative to solar and 
is remarkably close to 
the ratio of heating rates. Either this is a chance coincidence or
it means that $G_{0}$  in DLAs is nearly the same as in the 
ISM and 
the paucity of grains accounts for the lower
rate of heating in DLAs. Consequently
we shall adopt the grain photoelectric heating
mechanism.

\section{MULTI-PHASE STRUCTURE OF THE DLA GAS} 
\label{mdls}

To further evaluate the grain photoelectric heating hypothesis, we compute the thermal 
equilibrium temperature as a function of density for  gas
subjected to photoelectric grain heating. We adopt the treatment of 
W95 who calculate the two-phase structure of neutral
gas in the ISM. In this calculation the gas is assumed to be mainly
atomic and in a state of thermal and ionization equilibrium.  We also
include heating
and ionization due to cosmic rays, soft X-rays, and the photoionization of 
C I by FUV radiation (cf. W95).
Cooling is assumed to arise from fine-structure and
metastable transitions in ions of abundant elements, from 
{Ly$\alpha$}, and from radiative recombination of electrons onto
grains. Rather than repeat the
W95 analysis here, we summarize the important
points, emphasizing how the input physics for DLAs differs from that of the
ISM.

\subsection{Elemental Abundances and Dust-to-Gas Ratios}

Elemental abundances affect both the heating and cooling rates in DLAs
in different ways. Consider the heating rate.
The rate of grain photoelectric heating, {\gamdnr}, is proportional to the dust-to-gas
ratio in DLAs, $k_{DLA}$, 
which in turn depends on the fraction of extant metals in grains.
More specifically, {\gamdnr} depends on the grain composition, i.e.,
on whether grains in DLAs are mainly carbonaceous, as in the
Galaxy, or mainly silicates as in the SMC (see Weingartner \& Draine 2001b).
Recent evidence suggests that at $z$ $<$ 1, grains in DLAs are
Galactic. The unambiguous detection of the 2175 {\AA} graphite absorption 
feature and the overall shape of the reddening curves suggest that a known
DLA at $z$ = 0.524 (Junkkarinen {\etal} 2002) and a DLA detected in 
a gravitationally lensed galaxy at $z$ = 0.83 (Motta {\etal} 2002) contain
Galactic dust. On the other hand, a search for the 2175 {\AA} feature 
in 5 DLAs with mean redshift $\approx$ 2 (Pei {\etal} 1991) resulted in null detections
with upper limits on 2175 {\AA} optical depth significantly below predictions
based on the relative reddening of QSOs with foreground DLAs.
To account for this result, 
Pei {\etal} (1991) suggested that dust in DLAs resembles SMC dust
which does not exhibit the 2175 {\AA} feature, presumably because it
is composed mainly of silicates.

For these reasons we shall consider a ``Gal'' model in which
dust in DLAs consists of carbonaceous grains and PAHs and an ``SMC''
model in which the dust in DLAs consists of  silicate grains.
The ``Gal'' model assumes that small (4$-$10 {\AA}) carbonaceous
dust grains dominate the heating as they do in the Galaxy, and we
shall use the heating efficiency for Galactic regions rich in small
carbonaceous grains computed by Bakes {\&} Tielens (1994). We infer
the abundance of carbonaceous grains (per H atom) from the relative
dust-to-gas ratio, {\kapnr}, where {\kapnr}=$k_{DLA}/k_{MW}$ and
$k_{MW}$ is the dust-to-gas ratio in the Galaxy. We determine
{\kapnr} from the observed depletion of Fe in each DLA
(see derivation in the Appendix). Our method  assumes
that the number of C atoms depleted onto dust grains per depleted
Fe atom is the same in DLAs as in the Galaxy; i.e.,
\begin{equation}
{\Biggl(}{n^{C}_{depleted} \over n^{Fe}_{depleted}}{\Biggr)}_{DLA}={\Biggl(}{n^{C}_{depleted} \over n^{Fe}_{depleted}}{\Biggr)}_{MW} 
\label{eq:CtoFedepleted}
\perd
\end{equation}

\noindent This is a reasonable assertion because Fe tracks C in
metal-poor stars (Carretta {\etal} 2000). 
Indexing the C depletion to the Fe depletion is necessary because the C
abundance typically cannot be measured in DLAs and Fe is the only element
for which the depletion level is known (by comparison with Si).
We also assume the size distribution of the carbonaceous grains 
containing the depleted C atoms follows that of Bakes \& Tielens (1994).
As a result, the
grain photoelectric heating rate equals {\gamdnr} computed for pure carbonaceous
grains and solar composition multiplied by {\kapnr}
(equation 1).

The ``SMC'' model 
assumes that silicate grains dominate the
heating. The absence of a 2175\AA~ bump in the typical SMC dust extinction curve
and deficit of 12$\mu$m emission indicate a lack of small carbonaceous grains
(Sauvage \& Vigroux 1991),  
so it is the remaining 
small silicate grains that dominate the heating.  
Therefore, 
we use the heating efficiency 
calculated by Weingartner \& Draine (2001a) 
for regions that lack small carbonaceous 
grains. 
We infer the abundance of 
silicate grains from the dust-to-gas ratio determined 
from the observed depletion of iron analogous to the equation 
for carbon shown above. This assumes 
that the number of depleted Si atoms per depleted Fe atom 
is the same in DLAs as in the Galaxy, making the SMC 
model a hybrid of SMC and Galactic conditions.   
Observations of [Fe/Si] in DLAs will 
allow us to check this assumption against reality.  
In calculating the gaseous carbon 
abundance and using [Fe/Si] to determine $\kappa$, we assume 
that C and Si are nearly undepleted.  The possible contradiction between 
using the heating efficiency of carbonaceous or silicate grains  
while assuming C and Si to be undepleted will be discussed 
in $\S$ 6.

To estimate
{\kapnr} 
we compute 
the fraction of Fe in grains.
To estimate this fraction we need to determine the intrinsic abundance
of Fe and compare it to its gas-phase abundance. Although previous workers used
Zn to estimate intrinsic Fe, we shall use Si because
(1) 
the median ratio $<$[Si/Zn]$>$=0.03$\pm$0.05 for a sample of 
12 DLAs indicates Si traces Zn (implying that Zn may trace $\alpha$-enhanced
elements rather
than Fe-peak elements (PW02)), (2) there is only one case of
Si  depletion
 to date (Petitjean {\etal} 2002),
and (3) Si abundances have been measured out to $z$ = 4.5 whereas Zn is rarely
measured at $z$ $>$ 3.3. 
In the Appendix we show that in the case of grains composed of Fe,   
\begin{equation}
{\kappa} = 10^{[{\rm Si/H}]_{int}}{\Bigl (}10^{{\rm [Fe/Si]}_{int}}-10^{[\rm Fe/Si]_{gas}}{\Bigr )}
\label{eq:kappadef}
\perd
\end{equation}

\noindent In the last equation the abundance ratios
[X/Y]=log$_{10}(X/Y)$$ \\
-$log$_{10}(X/Y)_{\odot}$, and the subscripts ``int''
and ``gas''  refer to intrinsic and gas phase
abundance ratios.

Since {\kapnr} is sensitive to the level of depletion, 
we test two models. In the minimal depletion model  we 
make use of  abundance patterns of DLAs deduced by PW02.
Though they
derived the median value $<$[Fe/Si]$_{int}$$>$ = $-$0.3
from the abundance pattern of metal-poor DLAs
which are not expected to have significant dust depletion,
we shall adopt the more conservative value of $-$0.2.
This is consistent with {$\alpha$} enhancement expected from type II
supernovae which dominate nucleosynthesis at high $z$. We then  
let [Fe/Si]$_{gas}$ equal the observed ratio and
use equation ({\ref{eq:kappadef}}) to 
derive {\kapnr} for each DLA.
Because the value of  [Fe/Si]$_{int}$  is not yet well established, we consider 
a second, maximal depletion model in which we derive {\kapnr} by assuming [Fe/Si]$_{int}$=0;
i.e., we assume the observed deviations of the Fe to Si ratios 
from the solar value are caused only by
depletion.
In both models we assume
[Si/H]$_{int}$=[Si/H]$_{gas}$ because of evidence that Si is nearly undepleted. In cases
where only observational limits exist on [Si/H]$_{gas}$ or [Fe/H]$_{gas}$ we 
substitute elements such as S or Zn for Si, and Ni, Cr, or Al for Fe.
One could also consider the 
prescription of Pei {\etal} (1999) who let the dust-to-gas
ratio equal the observed Fe abundance, i.e.,  {\kapnr} = 10$^{{\rm [Fe/H]_{gas}}}$, 
but that turns out to be intermediate between the two models considered below; 
the minimal depletion model  yields the smallest values of {\kapnr} and the
maximal depletion model  yields the
largest values of {\kapnr}.

The abundances of the elements
C, O, Si, and Fe in the gas phase affect the cooling rate since 
transitions of C$^{+}$, O$^{0}$, Si$^{+}$, and Fe$^{+}$  are the major coolants. 
In our models we use the following prescription to compute gas phase
abundances of these elements in each DLA.  First we equate the
intrinsic abundance of Si to its measured gas-phase abundance; i.e.,
[Si/H]$_{int}$=[Si/H]$_{gas}$ where the latter are listed in Table 1.
Oxygen resembles Silicon as it is undepleted and an
{$\alpha$}-enhanced element. As a result we assume that
[O/H]$_{gas}$=[Si/H]$_{int}$. While we assume that carbon is
undepleted, we are aware this poses a potential contradiction with the
``Gal'' dust model in which the depletion level of C is assumed
proportional to the depletion level of Fe and we address this issue in
$\S$ 6. In any case C is not an {$\alpha$}-enhanced element, but
rather is likely to trace Fe since [C/Fe]$\approx$0 in stars of all
metallicities (e.g. Carretta {\etal} 2000).  Therefore we assume
[C/H]$_{gas}$=[Fe/H]$_{int}$ where
[Fe/H]$_{int}$=[Fe/Si]$_{int}$$+$[Si/H]$_{int}$.
For consistency with the other model abundances we compute
[Fe/H]$_{gas}$ from [Fe/H]$_{int}$ rather than equate it to the
observed DLA Fe abundance. In this case [Fe/H]$_{gas}$=[Fe/H]$_{int}$\\ 
$+$log$_{10}$[1-{\kapnr}{$\times${10$^{-[Fe/H]_{int}}$}}].  The
results are summarized in Table 2.

\begin{table}[h] 
\begin{center}
\caption{{\sc Gas-Phase Abundances}}
\begin{tabular}{lcc}
Element &log$_{10}$(X/H)$_{\odot}$ &log$_{10}$(X/H)$_{gas}$$-$log$_{10}$(X/H)$_{\odot}$   \\
\tableline
He & $-$1.00      &   0  \\
C & $-$3.44      &   [Si/H]$_{int}$$+$[Fe/Si]$_{int}$  \\
O & $-$3.34      &   [Si/H]$_{int}$  \\
Si & $-$4.45      &   [Si/H]$_{int}$  \\
Fe & $-$4.45      &   [Si/H]$_{int}$$+$[Fe/Si]$_{int}$ \\
   & &     $+$log$_{10}$[1-{\kapnr}{$\times$}10$^{-[Fe/H]_{int}}$]  \\

\end{tabular}
\end{center}
\end{table}

\subsection{Heating and Cooling}

In this subsection we discuss the sources of heating and cooling in
DLA gas. Specifically, we consider a gas layer
subjected to  heating by grain photoelectric emission, cosmic ray
ionization, X-ray ionization, and photoionization of C I. We also
discuss cooling by important emission lines from abundant elements,
and show how direct excitation by CMB photons
and indirect excitation due to pumping
by FUV fluorescence photons cause
the spontaneous emission rate to deviate from the cooling rate.
Similar discussions, that did not include optical pumping,
were given 
by Spaans \& Norman (1994) and subsequently by Liszt (2002).

\subsubsection{Heating}

The heating rate is given by
\begin{equation}
{\Gamma} = {\Gamma}_{d} + {\Gamma}_{CR} + {\Gamma}_{XR} + {\Gamma}_{{\rm CI}}
\label{eq:gamtot}
\cmma
\end{equation}

\noindent where $\Gamma_{d}$ is given by equation ({\ref{eq:Gamddef}}), and ${\Gamma}_{CR}, \ {\Gamma}_{XR}, \ {\rm and}
\  {\Gamma}_{{\rm CI}}$ are the heating rates due to cosmic rays, X-rays, and photoionization
of C I by the FUV radiation field, $G_{0}$. We ignore heating due to
the integrated background  from galaxies and QSOs as it is negligible
compared to {\gamdnr} for the range of SFRs considered in $\S$ 5; i.e., log$_{10}${\ps}
$>$ $-$4.0 {\smpykpc}.    
We compute $\Gamma$
by adopting expressions and parameters used by W95 to model the
ISM, but where appropriate extrapolate to physical conditions pertaining to DLAs.
Thus, in the case of ``Gal'' dust
we compute {\gamdnr} by adopting the Bakes \& Tielens (1994) expression
for the photoelectric efficiency,
since we assume that the DLAs have the same relative distribution of small grains and
PAHs as
the ISM.  In the case of ``SMC'' dust
we compute {\gamdnr} by adopting the Weingartner \& Draine (2001a)
expression for photoelectric efficiency in the case of pure silicates,
black-body FUV radiation,
and selective extinction $R_{v}$=3.1.
On the other hand there 
is no 
{\em a priori}
reason why $G_{0}$  in DLAs should equal 1.7, the 
widely accepted value for the ISM (Draine 1978).
As we show in $\S$ 5, $G_{0}$ $\propto$ {\ps} and the SFRs per
unit area in DLAs need not equal the Milky Way rates.
Moreover, the transfer of FUV radiation depends on
the dust optical depth which should be lower in DLAs than in the ISM.
We use the inferred optical to determine 
a self-consistent solution for {\ps}, which reveals
the SFR per unit area (see $\S$ 5.2).

To compute {${\Gamma}_{CR}$} we assume
{${\Gamma}_{CR}$}=$\zeta_{CR}$$E_{h}(E)$ where expressions for
$\zeta_{CR}$, the primary 
cosmic-ray ionization rate, and $E_{h}(E)$, the energy deposited for each
primary electron of energy $E$, are given by W95.
These authors
find $\zeta_{CR}$ = 1.8$\times$10$^{-17}$ s$^{-1}$ in the Galaxy.
We scale this result to DLAs by assuming
\begin{equation}
{\zeta_{CR}} = 1.8{\times}10^{-17}{\Biggl (}{\dot{\psi_{*}} \over 10^{-2.4} {\rm M_{\odot} yr^{-1} kpc^{-2}}}{\Biggr )} s^{-1}
\label{eq:zetaCR}
\cmma
\end{equation}

\noindent where we have used log$_{10}${\ps} = $-$2.4 {\smpykpc} for the disk of the Galaxy 
(Kennicutt 1998).

To compute the effects of soft X-rays
we use the W95 expressions for the heating rate, $\Gamma_{XR}$, and 
primary and total ionization rates, $\zeta_{XR}$
and $\xi_{XR}$. We again scale to DLAs by assuming all these quantities
are proportional to {\ps}. W95 assume soft X-rays (photon energies
exceeding 0.2 keV) are emitted by thermal and non-thermal components.
The thermal component comprises the hot ($T$ $\sim$ 10$^{6}$ K) 
coronal phase of the ISM. The non-thermal component consists of
extragalactic power-law radiation. These X-rays penetrate
the outsides of CNM clouds, heating the gas to form the
WNM (e.g. Heiles 2001). W95 assume the incident X-rays are attenuated
by a WNM layer of gas with H I column density log$_{10}$$N_{w}$ 
= 19 cm$^{-2}$. For our analysis we shall assume 
log$_{10}$$N_{w}$ = 20 cm$^{-2}$ instead. This is because
low-density ($n$ $\sim$ 0.1 cm$^{-3}$) WNM gas cannot remain
neutral at H I column densities log$_{10}$$N_{w}$ $<$ 20.3 
cm$^{-2}$ due to the background ionizing radiation field which 
is about 100 times more intense at high redshifts than at $z$ = 0
(e.g. Prochaska \& Wolfe 1996).
Had we assumed log$_{10}$$N_{w}$
=19.0 cm$^{-2}$, the total column density of neutral plus
ionized gas would exceed log$N$=20.0 cm$^{-2}$, and it
is the total column density which is crucial for determining
X-ray opacity. 
Furthermore, Vladilo {\etal} (2001) present
strong arguments that gas in DLAs is mostly neutral. 
Thus our limit is conservative. Another reason for adopting
the larger WNM column density is our model assumption that a significant
fraction of the H I column density in each DLA consists of WNM gas. 
The result of these model assumptions is that
cosmic rays will be a more important source of heating
and ionization than X-rays.

Notice, that we have assumed that $G_{0}$, $\Gamma_{CR}$, and $\Gamma_{XR}$ are all proportional
to {\ps}. While unproven for $\Gamma_{CR}$ and $\Gamma_{XR}$, we believe this assumption is reasonable as all
three quantities are ultimately driven by the formation rate of massive stars. 
This is because cosmic rays are thought to be accelerated in supernova remnants, and
much of the soft X-ray emission is thought to arise in hot gas located behind supernova
shocks (McKee \& Ostriker 1977).

\subsubsection{Cooling}

The cooling rate (erg cm$^{3}$ s$^{-1}$) is given by
\begin{equation}
{\Lambda} = {\Lambda}_{FS} + {\Lambda}_{MS} +
{\Lambda}_{Ly{\alpha}}+{\Lambda}_{GR}
\label{eq:Lamtot}
\perd
\end{equation}
\noindent At $T$$<$3000 K,
cooling is dominated by the fine-structure term, $\Lambda_{FS}$. 
The leading contributors are emission by
the fine-structure lines [C II]
158 {\micron}, which typically dominates at
$T$ $<$ 300 K, and [O I] 63 {\micron} which is comparable to
158 {\micron} emission only at $T$ $>$ 300 K. Following W95
we also include fine structure cooling from other transitions
in O$^{0}$ (i.e., neutral Oxygen) and from
transitions in Si$^{+}$  and Fe$^{+}$. At $T$$>$ 3000 K,
the term $\Lambda_{MS}$ becomes important. This arises from  excitation of 
metastable transitions of C$^{+}$, O$^{0}$, Si$^{+}$, and S$^{+}$.  At
higher temperatures, the {\lya} cooling term, $\Lambda_{Ly{\alpha}}$ starts
to dominate along with $\Lambda_{GR}$, the grain recombination rate. 
We computed  $\Lambda_{GR}$ by adopting
the Bakes \& Tielens (1994) expression for cooling
due to radiative recombinations of electrons onto PAHs and grains. 
Note, we have not included cooling by transitions in the neutral
species C$^{0}$, Fe$^{0}$, Mg$^{0}$, and Si$^{0}$ considered by
W95 as their contribution to $\Lambda$ is negligible.

By definition the cooling rate of the gas equals the net loss of thermal kinetic energy
per unit time. In the case of line cooling
this is the product of (a) the difference between
the collisional excitation and de-excitation rates, and (b) the
energy of the atomic transition.
The former equals
the spontaneous emission rate, provided collisions are the dominant source of
excitation and de-excitation. As a result $n{\Lambda}_{{\rm CII}}$
= {\lcrnr} in the ISM
where {\lcrnr} is the spontaneous emission rate per H atom 
of the $^{2}P_{3/2}$ $\rightarrow$$^{2}P_{1/2}$ transition.
However, this equality can break down in DLAs since radiative excitation can be
important. At high $z$ the CMB contributes significantly to
the rate at which the 
$^{2}P_{1/2}$ and $^{2}P_{3/2}$ states are populated. Moreover, for large values of
$G_{0}$, these ground-term fine-structure
states can be populated indirectly through FUV excitation of higher energy states;
i.e., through optical pumping 
(termed ``fluorescence'' by Silva \& Viegas 2002). When radiative
excitations are important we have 
\begin{equation}
\begin{array}{ll}
l_{cr}= n{\Lambda}_{\rm CII}+(l_{cr})_{pump}+(l_{cr})_{CMB}    
\cmma
\end{array}
\label{eq:lcall}
\end{equation}

\noindent where 
({\lcrnr})$_{pump}$ and ({\lcrnr})$_{CMB}$
are the spontaneous energy emission rates in the limits of
pure optical pumping and CMB excitation. In deriving the last
equation we used the condition
1+z $<<$$h{\nu}_{ul}/kT_{CMB}$=33 where $h{\nu}_{ul}$
is the excitation energy corresponding to the $^{2}P_{3/2}$$\rightarrow$
$^{2}P_{1/2}$ transition in C II, and $T_{CMB}$=2.78 K, the current
temperature of the CMB. We find that

\begin{equation}
\begin{array}{ll}
(l_{cr})_{pump}=({\rm C/H}){\Gamma}_{lu}h{\nu}_{ul} \ , \\   
(l_{cr})_{CMB}=2({\rm C/H})A_{ul}h{\nu}_{ul}{\rm exp}{\biggl [}-h{\nu}_{ul}/{\bigl [}k(1+z)T_{CMB}{\bigr ]}{\biggr ]} \
\cmma
\end{array}
\label{eq:lcpumpcmb}
\end{equation}

\noindent where (C/H) is the carbon abundance, and
$A_{ul}$ is the rate of 
spontaneous emission for the 
$^{2}P_{1/2}$
$\rightarrow$ $^{2}P_{3/2}$ transition.
The quantity ${\Gamma}_{lu}$ is the {\em  net} rate at which
state $l$ pumps state $u$ 
(see Silva \& Viegas 2002). 
We calculated {\lcrnr} using standard expressions for excitations due to
collisions and CMB radiation. We used the 
Silva \& Viegas (2002) code, POPRATIO, to compute 
the rate at which the
$^{2}P_{1/2}$ and $^{2}P_{3/2}$ states are populated
by optical pumping. We considered  
indirect excitation of the $^{2}P_{3/2}$ and $^{2}P_{1/2}$ states
through transitions to 8  higher levels. 
For consistency we used the
spectral form advocated by Draine (1978) for the FUV radiation field,
normalized such that $G_{0}$=$\int 4{\pi}J_{\nu}d{\nu}$/(1.6${\times}10^{-3}$
ergs cm$^{-2}$ s$^{-1}$). However, this procedure ignores the effects
of line opacity, which can effectively suppress the pumping rate
when the gas is optically thick to transitions such as C II 1334.5, 
and CII 1036.3 
(Sarazin {\etal} 1979; 
Flannery {\etal} 1979; 1980). Because the values of $G_{0}$,
resonance-line optical depth, and collisional excitation
rates are required to evaluate the suppression of optical pumping,
we shall re-evaluate our ``optically thin'' solutions 
for self-consistency in $\S$ 5.2.2.

\subsection{Phase Diagrams for DLAs}

In this subsection we compute the two-phase structure of neutral gas in DLAs.
We find that if the gas is CNM, the SFRs per unit area
deduced for local disk galaxies (Kennicutt 1998)  
generate [C II] spontaneous emission rates similar to those
observed
in DLAs  for typical DLA metallicities.
If the gas is WNM, the required SFR per unit area must be 
{\em  at least} a factor of 10 higher.

We solved the equations of thermal and ionization equilibrium for gas at
constant density with 
the numerical  techniques and iterative procedures outlined in W95.
We checked our technique by computing
solutions for ISM conditions. In that case we assumed $G_{0}$=1.7, [Si/H]$_{int}$=0,
log$_{10}${\kapnr}=[Fe/H], and the same density-dependent depletion formulae
advocated by W95. The results are in good, though
not exact, agreement. Most importantly, the {\lcrnr} versus $n$ curves
are in excellent agreement with the W95 results except at
log$_{10}$$n$ $<$ $-$0.5 cm$^{-3}$ where  optical pumping effects ignored by
these authors
causes {\lcrnr} to deviate significantly above $n{\Lambda_{{\rm CII}}}$.

\begin{figure*}[ht]
\scalebox{0.65}[0.6]{\rotatebox{-90}{\includegraphics{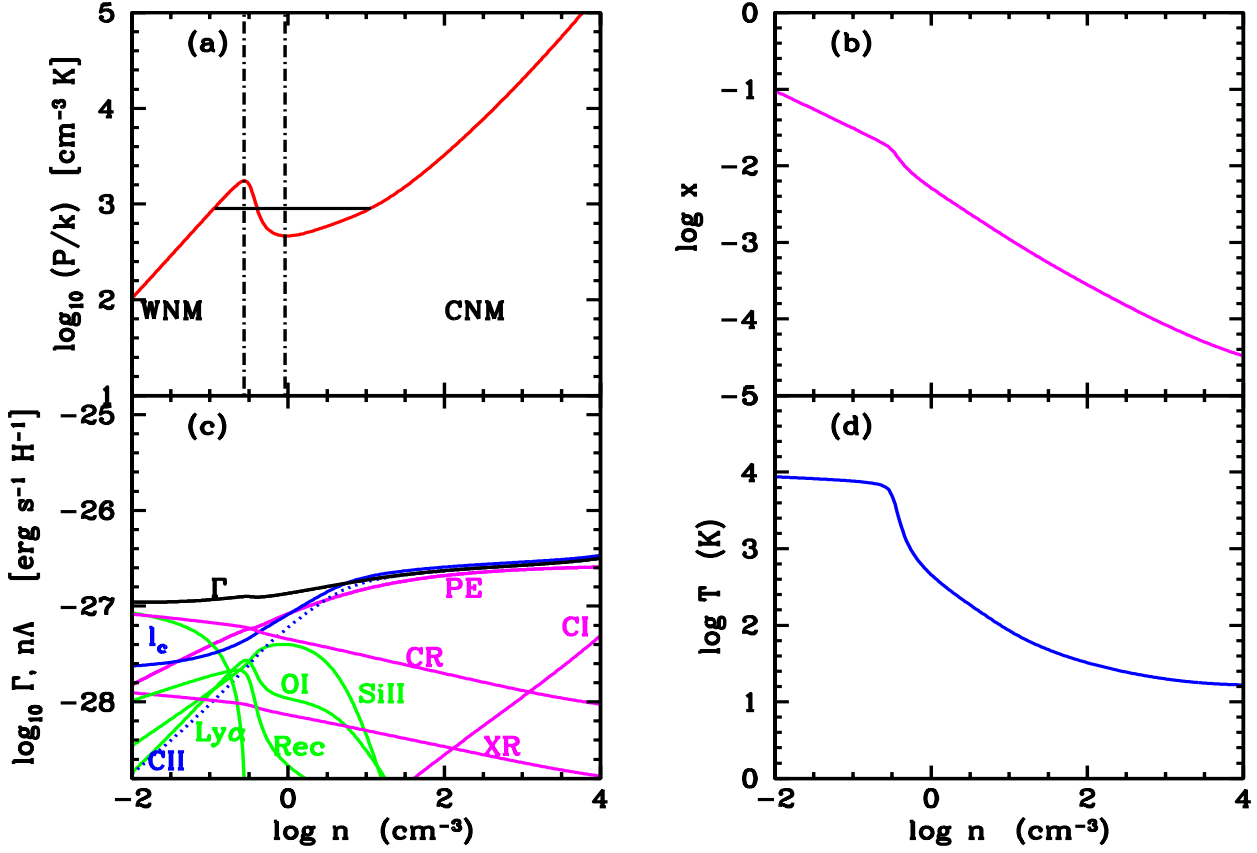}}}
\caption{Two-phase diagrams for gas heated by grain photoelectric emission
plus cosmic rays and soft X-rays, where  the SFR per unit area, log$_{10}${\ps}
=$-$2.4 {\smpykpc}. Metallicity, [C/H]=$-$1.5 and dust-to-gas ratio
log$_{10}$$\kappa$=$-$1.7. Panel (a) shows pressure versus density. The
S shape curve is indicative of a two-phase medium.
Labels and horizontal line in ($n,P$) plane explained in 
text. Panel (b) shows fractional
ionization versus density. Magenta curves in panel (c) show grain photoelectric heating (PE),
cosmic-ray heating (CR), 
X-ray heating (XR), and CI photoionization heating rate
versus density. Dotted blue curve is [C II] 158 {\micron} cooling rate, and 
green curves are [O I], [Si II], {\lya}, and 
grain recombination cooling rates. Solid blue curve
is [C II] 158 {\micron} spontaneous energy emission rate.
Black curve ($\Gamma$) is total heating rate.
Panel (d) shows temperature versus density. }
\label{twophase}
\end{figure*}

To illustrate the behavior of two-phase media with DLA 
conditions we let [Si/H]$_{int}$ = $-$1.3, the mean 
Si abundance found for 
DLAs (PW02). We assume the ``Gal'' model and use maximum depletion to find
log$_{10}${\kapnr} = $-$1.5. We assume an ISM
radiation field, G$_{0}$ = 1.7, and adopt the mean redshift
of the DLA sample, $z$ =2.8, to compute the CMB temperature. We compute the 
cosmic ray  and X-ray heating rates from equation ({\ref{eq:zetaCR}}) by assuming the ISM
star formation rate
log$_{10}${\ps}=$-$2.4 $M_{\odot}$ kpc$^{-2}$ yr$^{-1}$.  
The resulting equilibrium  curves shown in Figure 3 
exhibit 
the same two-phase
equilibria found by W95 for the ISM. 
In a  plot of pressure, $P/k$, versus density, $n$, (see Fig. 3a) the regions
of thermal stability occur where 
$\partial(logP)$/$\partial (logn)$ $>$ 0 (in the case
of constant $\Gamma$).  Thus a two-phase medium
in which a WNM can remain in pressure equilibrium with a CNM
can be maintained between $P^{min}/k$ $\approx$ 460 K cm$^{-3}$
and $P^{max}/k$ $\approx$ 1750 K cm$^{-3}$. An example in which
$P=(P_{min}P_{max})^{1/2}$ is shown as the horizontal line
connecting the WNM and CNM. The intercepts with the $P(n)$ curve
in the WNM and CNM correspond to thermally stable states;
a WNM 
with $T$ $\approx$ 7600 K 
and log$_{10}$$n$ $\approx$ $-$ 1 cm$^{-3}$ in
presure equilibrium with a CNM with $T$ $\approx$ 80 K
and 
log$_{10}$$n$ $\approx$ $+$1 cm$^{-3}$. Gas with 
densities $-$ 0.6 $<$ log$_{10} n$ $<$ 0.0 cm$^{-3}$ is thermally unstable and
evolves either to WNM or CNM states.
Figure 3b shows
the fractional ionization as a function of density.

Figure 3c plots the
heating rates, $\Gamma$ (red curves), cooling rates, 
$n{\Lambda}$ (green curvesand dotted blue curve in the case of C II),
and the spontaneous emission rate {\lcrnr} (solid blue curve).
It is evident that grain photoelectric heating dominates
in the CNM while cosmic ray heating dominates in the WNM 
(see W95). By contrast to the ISM,  cosmic rays dominate
X-ray heating in DLAs
for all densities, owing to the higher X-ray opacity
of the H I column density assumed for DLAs.
The dominant coolant in the CNM is  
[C II] 158 {\micron} radiation  which 
is insensitive to
density at 0.5 $<$  log$_{10} n$ $<$ 4.0 cm$^{-3}$. This breaks down at
log$_{10}$$n$ $>$ 4.5 cm$^{-3}$ (not shown) where C I photoionization dominates
the heating rate and the cooling rate increases rapidly with
density. On the other hand [C II] 158 {\micron}
emission comprises less than 10$\%$ of the
cooling in the WNM. Furthermore owing to CMB excitation
and optical pumping,
the population of the $^{2}$P$_{3/2}$ state of
C$^{+}$ is larger than in the case of collisional 
excitations and de-excitations
alone. 
As a result, 
at log$_{10} n$ $<$  0.0 cm$^{-3}$
the spontaneous emission rate, {\lcrnr},
will exceed the cooling 
through 158 {\micron} emission (see equation {\ref{eq:lcall}}).
Notice that {\lcrnr} is
insensitive to density in the CNM where {\lcrnr}$\rightarrow$
{\gamdnr} and in the WNM where {\lcrnr} ${\rightarrow}$ ({\lcrnr})$_{pump}$
(see equation {\ref{eq:lcpumpcmb}}).
This reduces the 
uncertainty in estimating {\lcrnr} in the models discussed in $\S$ 5.
In any case the spontaneous emission rate in the CNM, 
log$_{10}$ {\lcrnr} $\approx$ $-$26.6 ergs s$^{-1}$ H$^{-1}$, 
is comparable to the mean {\lclos} for the DLA sample in Figure 1.
Stated differently, the hypothesis of grain photoelectric heating 
can account for the mean [C II]  cooling rate of DLAs provided
(a) heating occurs in CNM gas with a low dust-to-gas ratio, and (b)
heating is driven by an FUV radiation field  
comparable to that inferred for the ISM.
On the other hand the observed {\lclos} could also arise in 
WNM gas (cf. Norman \& Spaans 1997; Kanekar and Chengalur 2001), 
provided $G_{0}$ 
is about a factor of 30 or more higher. We will
address this further in  $\S$5.

In our models
the heating rates for DLAs were inferred from the cooling rates by
assuming steady-state conditions. To determine whether this assumption is valid,
consider the cooling time of gas in pressure equilibrium,
\begin{equation}
t_{cool} = {{(5/2)(1.1 + x)kT} \over {n{\Lambda}}}
\cmma
\end{equation}

\noindent (see equation 10 in W95). We find that
$n{\Lambda}$=3{$\times$}10$^{-27}$ ergs s$^{-1}$ H$^{-1}$ is required for CNM gas to
match the inferred [C II] emission rate. Since $T$$\approx$ 50 K, 
we have $t_{cool}$ $\approx$
3{$\times$}10$^{5}$ yrs. If the DLA gas is in the WNM phase, then   
$n{\Lambda}$=3{$\times$}10$^{-26}$ ergs s$^{-1}$ H$^{-1}$
is required to explain the observed {\lclos}. In that case $T$ $\approx$
8000 K and $t_{cool}$ $\approx$ 5{$\times$}10$^{6}$ yrs. Because these time-scales
are comparable to the dynamical time-scales of individual interstellar clouds, the
assumption of thermal balance may break down on the spatial scales of
typical interstellar clouds. However,
the measured quantity for DLAs is {\lclos} which is the density-weighted average
of the {\ciis} spontaneous emission
rate, {\lcr}, along the sightline through a typical DLA. In this case the
relevant dynamical
time scale is that of a typical protogalaxy which in any scenario is large compared
to 
5{$\times$}10$^{6}$ yrs. Stated differently the fluctuations of heating and cooling
rates integrated over the length scales of typical DLAs average out so that
the mean rates are equal. As a result, the assumption of thermal and ionization balance
should be
an excellent approximation for DLAs.

On the other hand the assumption of pressure equilibrium 
is not well established empirically.
Accurate Arecibo 21 cm measurements of H I spin temperatures in the ISM
reveal strong evidence for CNM gas with $T$=25 $\rightarrow$ 50 K, but no 
evidence for WNM gas with $T$ $>$ 7000 K. Rather, a significant
fraction of the warm gas lies in the thermally unstable regime with
$T$=500 $\rightarrow$ 5000 K (Heiles 2001). While these measurements
do not rule out multi-phase models for the ISM, 
they
bring to mind alternative scenarios. Specifically, Vazquez-Semadeni
{\etal} (2000; see also Gazol {\etal} 2001) compute 2D numerical
simulations in which
the dynamics of ISM clouds are dominated by turbulence rather
than thermal instability. In this scenario the boundaries of dense
CNM clouds are accretion shocks comprising thermally unstable gas
rather than quiescent contact discontinuities separating disparate
phases at constant pressure. Moreover the unstable
gas is found to comprise a significant fraction of the ISM mass (although,
higher-resolution 3D numerical
simulations by Kritsuk \& Norman (2002) find the fraction to 
be less than 15$\%$).
But, as stressed by Vazquez-Semadeni
{\etal} (2000), the cooling times in the thermally unstable
gas are shorter than the dynamical time scales, and as a result
the thermally unstable gas evolves quasi-statically through a 
sequence of 
thermal equilibrium states. 
For these reasons, the cooling curves shown in Figure 3c
also apply to the ``turbulence scenarios''.  Although
$n$${{\Lambda}_{{\rm CII}}}$ and {\lclos} increase as
$T$ decreases from 7500 K to 1000K, both quantities are
still small compared to the total heating rate, $\Gamma$.
As a result, relatively large SFRs are required for scenarios 
in which {\ciis} absorption occurs in warm neutral gas
(as discussed in Paper II), whether
or not that gas is in pressure equilibrium with the CNM.
And it is this property which ultimately rules out the WNM models.
We conclude that the possible break down of pressure equilibrium
has little effect on our results.

\section{THE STAR FORMATION RATE PER UNIT AREA}

We now estimate the SFR per unit area 
for each of
our sample of DLAs. 
We first solve the transfer equation 
for
the mean intensity of FUV radiation corresponding to the
{\kapnr} derived for each DLA and for a wide range of
{\ps}. For each DLA we assign appropriate gas-phase
abundances and then combine equation ({\ref{eq:Gamddef}}) with the steady state assumption
of equation ({\ref{eq:Gameqnlam}})
to compute {\lcrnr}.
We compare the computed {\lcrnr}
with the observed
{\lclos} to deduce {\ps} for each DLA in both the CNM
and WNM models. As we shall show, these SFRs are global in
nature as they correspond to {\ps} averaged over the
entire DLA.

\subsection{Solutions to the Transfer Equation}

Assume 
the gas, dust and stars comprising DLAs are {\em uniformly} distributed
throughout  plane-parallel disks with 
half-width, $h$, and radius, $R$.
A disk is a reasonable approximation for DLAs because dissipative collapse of
gas in all galaxy formation scenarios, including protogalactic
clump formation predicted by CDM numerical simulations (Haehnelt
{\etal} 1998), occurs along a preferred
axis, resulting in configurations resembling
plane-parallel layers. Though uniformity is a highly idealized assumption,
we shall show that the results do not differ qualitatively when
this assumption is relaxed (see Paper II).
We compute
the mean intensity, $J_{\nu}$, at the center and midplane location of the uniform disk 
and find:
\begin{eqnarray}
 J_\nu  = \frac{2}{4 \pi} \int_0^{2 \pi} d{\phi} {\Biggl [}  
\int_0^{\theta_c} \int_0^{h \sec \theta} r^2 \sin \theta d \theta dr 
\rho_\nu \frac{e^{-k_{\nu}r}}{4 \pi r^2} + \nonumber\\
\int_{\theta_c}^{\pi/2} \int_0^{R \csc \theta} r^2 \sin \theta d \theta dr 
\rho_\nu \frac{e^{-k_{\nu}r}}{4 \pi r^2} 
{\Biggr ]}      ,
\label{eq:Jnu_disk}
\end{eqnarray}

\noindent where
we have ignored scattering of photons, and 
$k_{\nu}$ is the absorption opacity of dust at frequency $\nu$.
The quantity  $\rho_\nu$ is the luminosity density of the uniform disk, 
$\theta_{c}$=$tan^{-1}(R/h)$,  
and the extra factor of two out front comes from having $\theta$ run 
from 0 to $\pi/2$.
After integration
we find that

\begin{eqnarray}
J_{\nu}={1 \over 2}{({\Sigma_{\nu}/4{\pi})} \over h{k_{\nu}}}{\Biggl[}1-{\Biggl (}{h \over {\sqrt{h^{2}+R^{2}}}}{\Biggr)}{\rm exp}(-k_{\nu}R)-\nonumber\\
{\int_{1}^{{\sqrt{h^{2}+R^{2}}/h}}{dx \over x^{2}}{\rm exp}(-k_{\nu}hx)}{\Biggr]}
\label{eq:Jnueqexp}
\perd
\end{eqnarray}
\noindent Note, we obtained 
equation (15) by assuming the radial distance to the edge
of the disk
equals $R$ for $\theta_{c}$$<$$\theta$$<$$\pi$/2.

To compute $J_{\nu}$ it is necessary to evaluate the quantities
$k_{\nu}$$h$ and $k_{\nu}$$R$; i.e., the
dust optical depths perpendicular and parallel to the plane of the disk.
Define the optical depth, ${\tau}_{\nu}$, to be that of an average line-of-sight through the disk.
At an average inclination angle of 45$^{o}$ we find that  
\begin{equation}
k_{\nu}{h}= {{\tau_{\nu}} / (2 {\sqrt2}})
\label{eq:knudef}
\perd
\end{equation}

\noindent To compute  {$\tau_{\nu}$}, we follow Fall \& Pei (1989) who
derived the following expression: 
\begin{equation}
\tau_{\nu}= [{A({\lambda})/ A(4400{\rm {\AA}})}]{k_{DLA}}[{N({\rm H I})/10^{21} {\rm cm^{-2}}}]
\label{eq:taunu}
\cmma
\end{equation}

\noindent where
$k_{DLA}$={\kapnr}{$\times$}$k_{MW}$ for 
``Gal'' dust and \\ 
$k_{DLA}$={\kapnr}{$\times$} 
[$k_{SMC}$$\times$10$^{-[Si/H]_{SMC}}$] for
``SMC'' dust,
and $A({\lambda})$ is the extinction at
wavelength {$\lambda$}=$c/{\nu}$.
The [Si/H]$_{SMC}$ term is the silicon abundance of the
SMC and appears because {\kapnr} is
normalized with respect to Galactic dust.
The photons 
responsible for photoelectric grain heating in DLAs have energies between
{$\cal E$} = 6 eV and the Lyman limit cutoff at 13.6 eV. At the characteristic
energy {$\cal E$} = 8 eV (corresponding to {$\lambda$}=1500 {\AA})
we  
find that
$k_{MW}$ and $A(1500{\rm {\AA}})/A(4400{\rm {\AA}})$ equal 0.79 and  2.5 for the ``Gal''
model,  and
$k_{SMC}$ and $A(1500{\rm {\AA}})/A(4400{\rm {\AA}})$ equal 0.05 and  5.0 for the ``SMC'' model.
We determine {$\tau_{\nu}$} by assigning
the appropriate {\kapnr} for each DLA, and by using the median of our sample
distribution of H I column densities, {\NH}=0.48{$\times$}10$^{21}$cm$^{-2}$.
For each DLA
we use the sample median rather than the measured value of {\NH}  because
{\NH} along a single line-of-sight is unlikely to represent the  H I
column encountered by most of the FUV radiation; since
{$\tau_{\nu}$} $<<$ 1, 
the FUV radiation which heats the grains is transported across the entire
DLA.
Consequently, {$\tau_{\nu}$} $\approx$ 0.01
for values of {\kapnr} typifying our sample.
\footnote{We use
equation ({\ref{eq:taunu}}) to determine {$\tau_{\nu}$} by adopting
the expression in equation ({\ref{eq:kappadef}}) for {\kapnr} to infer $k_{DLA}$
from either $k_{MW}$ or $k_{SMC}$. Though the definition of these dust-to-gas ratios
differs from that used to compute {\gamdnr}, the absolute values of $k_{DLA}$ are used only
for evaluating {\taunu}. To compute {\gamdnr}, all we need is the
relative quantity, {\kapnr}. Since we employ the same expression for {\kapnr}
for both purposes, we are assuming that all the intrinsic properties of the
grains are the same in DLAs and in present-day objects.}

These estimates of {$\tau_{\nu}$} imply the condition $k_{\nu}{h}$ $<<$ 1 in
every case. 
For reasonable aspect ratios, $R/h$, we find that $k_{\nu}{R}$ $<<$ 1 in most DLAs, but
in metal-rich objects the condition $k_{\nu}{R}$ $>$ 1 may hold.
Therefore, in the limits corresponding to most models, {\jnu} takes on the
following simple form:

\begin{equation}
J_{\nu} = {1 \over 2}{\Biggl (}{\Sigma_{\nu}/4{\pi}}{\Biggr )}\left\{
\begin{array}{ll}
{\biggl[}1+{\ln}(R/h)-{k_{\nu}}R{\biggr]}+O(k_{\nu}R)^{2}... \ ; \\
 {k_{\nu}}h \ << {k_{\nu}}R \ << 1 \\
{\biggl[}1-{\gamma}-{\ln}(k_{\nu}h)+0.5{k_{\nu}}h{\biggr]}+O(k_{\nu}h)^{2} \\
... ;{k_{\nu}}h \ << 1, {k_{\nu}}R \ >> 1
\cmma
\end{array}
\right.
\label{eq:Jnuapprox}
\end{equation}
\noindent where $\Sigma_{\nu}$ (=2$\rho_{\nu}$$h$) is the luminosity
density projected along the rotation axis of the disk,i.e., the
luminosity per unit area, and $\gamma$ is Euler's constant.  Equation
({\ref{eq:Jnuapprox}}) shows the mean intensity to vary linearly with
the source luminosity per area, but to be weakly dependent on
metallicity, reddening curve, and aspect ratio.  To illustrate the
parameter dependence of {\jnu} we solved equation (15) for (a) the
``Gal'' and ``SMC'' dust assumptions, (b) the maximal and minimal
depletion models, and (c) aspect ratios in the range prediced by
current models of galaxy formation. The results are shown in Figure 4
where we plot mean intensity versus [Si/H].  As expected, for fixed
values of $R/h$, {\jnu} increases with decreasing [Si/H] until the
disk becomes optically thin in the radial direction.  At lower values
of [Si/H], {\jnu} equals a constant that increases with increasing
$R/h$. At higher values of [Si/H], {\jnu} decreases with increasing
[Si/H], becoming insensitive to $R/h$ as the disk becomes optically
thick: this is because {\jnu} $\propto$ $\Sigma_{\nu}$/$\tau_{\nu}$ in
the optically thick limit.  Figure 4 also plots results for a uniform
sphere which, while unrealistic, provide a lower limit to the strength
of the radiation field.  For our standard model we shall adopt a disk
with an aspect ratio {$R/h$}=10. From Figure 4 we see that for a
measured [Si/H] the uncertainties in $R/h$ and dust composition cause
uncertainties in {\jnu} for the disk models which do not exceed $\sim$
30$\%$. When the spherical model is included these remain less than
50$\%$.

The grain heating rate is determined by $J$, i.e., {\jnu} integrated between
photon energies
{$\cal E$}= 6 and 13.6 ev. Assuming the Draine (1978) UV spectrum we find
that {\jnu} = 10$^{-19}$ ergs cm$^{-2}$ s$^{-1}$ sr$^{-1}$ Hz$^{-1}$
at {$\cal E$} = 8 eV (i.e., 1500 {\AA}) results in
4{$\pi$}{$J$}=1.6{$\times$}10$^{-3}$ ergs cm$^{-2}$s$^{-1}$,
which equals Habing's (1968) estimate of the UV interstellar radiation 

\begin{figure}[ht]
\centering
\scalebox{0.3}[0.3]{\rotatebox{-90}{\includegraphics{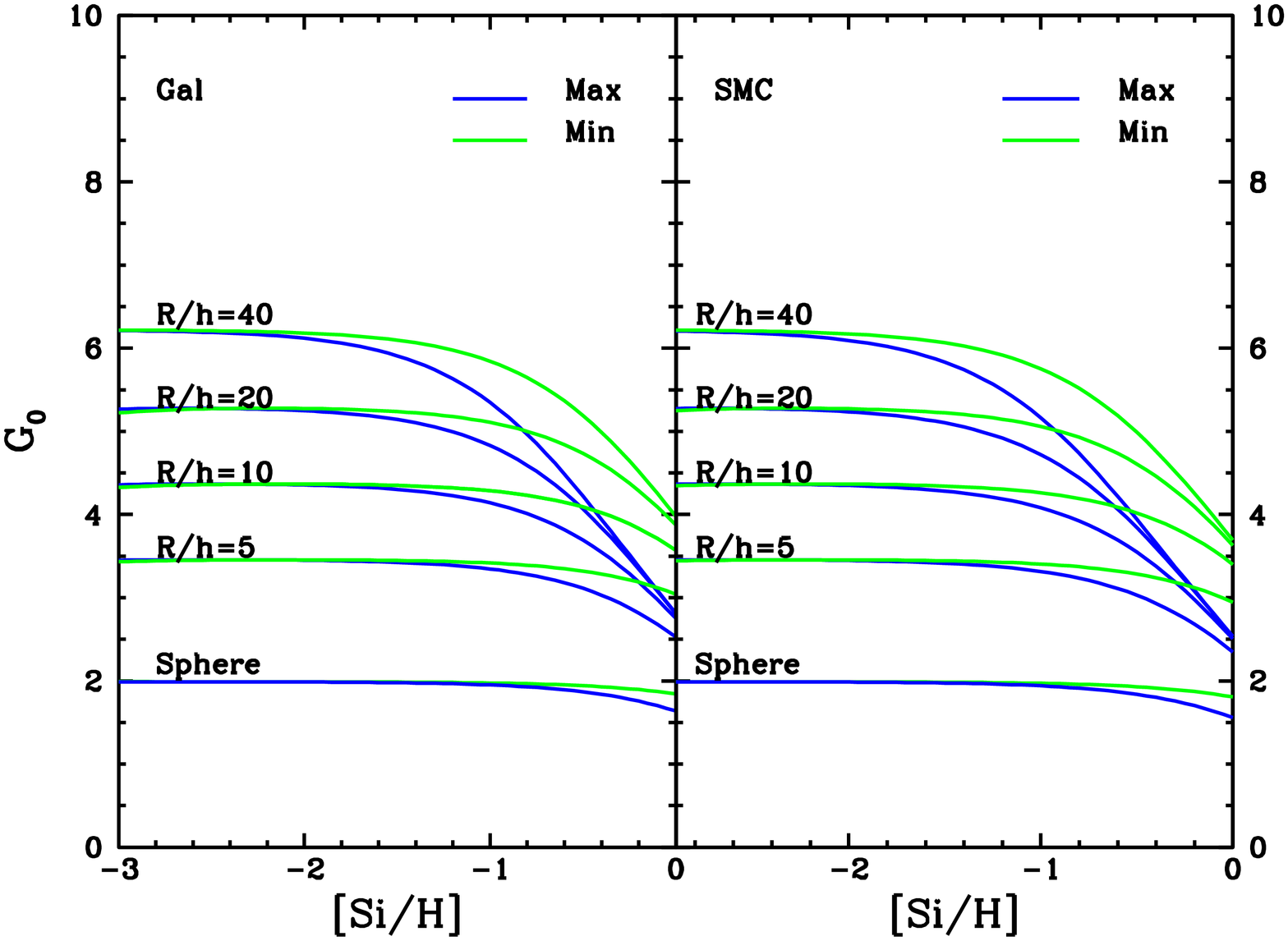}}}
\caption[]{Solutions to transfer equation given in equation (15).
The resultant
mean intensity $J_{\nu}$ (in units of
10$^{-19}$ ergs cm$^{-2}$ s$^{-1}$ sr$^{-1}$ Hz$^{-1}$),
which we denote as $G_{0}$, 
is plotted versus [Si/H] assuming
$\Sigma_{\nu}$ corresponding to a SFR, 
log$_{10}${\ps}=$-$2.4{\smpykpc}. In Figure 4a we assume ``Gal'' dust
and show results for various aspect ratios, $R/h$, and for the maximal
and minimal dust depletion models.  Figure 4b is the same as 4a, except that
``SMC'' dust is assumed. Spherical solutions also shown.}
\label{G0vskap}
\end{figure}

\noindent field. Therefore, we shall assume
\begin{equation}
G_{0}={\Biggl(}{J_{\nu} \over {1{\times}10^{-19} {\rm \ ergs \ cm^{-2} \ s^{-1} \ sr^{-1} \ Hz^{-1}}}}{\Biggr)}, \  {\lambda} = 1500 {\rm {\AA}}
\label{eq:G0}
\perd
\end{equation}

\noindent To relate 
$G_{0}$ to the rate of star formation we use the Madau \& Pozzetti (2000) calibration.
In that case 
\begin{equation}
{\Sigma_{\nu}}=8.4{\times}10^{-16}({\dot{\psi_{*}}}/{\rm \ M_{\odot} \  yr^{-1} \  kpc^{-2}) \  ergs \ cm^{-2} \ s^{-1} \ Hz^{-1}}
\label{eq:Signu}
\perd
\end{equation}

\noindent By combining equations (15), ({\ref{eq:G0}}) and
({\ref{eq:Signu}}), we can convert {$\dot{\psi_{*}}$} into $G_{0}$. To
see whether our technique reproduces the ISM radiation field, we
solved for $G_{0}$ assuming $R$= 20 kpc, $h$ = 0.125 kpc,
log$_{10}${\kapnr}=0, and log$_{10}${\ps} = $-$2.4 {\smpykpc}. We
found that $G_{0}$ = 1.6, which is in excellent agreement with the
Draine (1978) value of $G_{0}$ = 1.7.

Before changing topics we wish to emphasize several points.
First,  for typical metallicities the DLAs will
be optically thin to FUV radiation in every direction.
In the uniform disk approximation, sources at all distances contribute
roughly equally to {\jnu}. As a result, the {\em  SFRs  per unit area inferred from the
{\ciis} absorption profiles are representative of stars
distributed throughout the entire DLA, not just
in regions adjacent to the line of sight}.
This is in contrast to the ISM where the high dust opacity results 
in SFRs with only local
significance.  
Second, we computed {\jnu} for midplane points displaced from the
center of the disk; i.e., we computed {\jnu}($r,Z$) at cylindrical
coordinates ($r$,0). When $R/h$=10 we found {\jnu}($r,0$)/{\jnu}(0,0)
to slowly decrease from 1 at $r$=0 and to equal 0.9 at the median
radius $r=R$/{$\sqrt 2$}, 0.75 at $r=0.9R$, and 0.5 at $r=0.98R$; when
$R/h$=100, {\jnu}($r,0$)/{\jnu}(0,0)=0.94 at $r$=$R$/{$\sqrt 2$}, 0.85
at $r$=0.9$R$, and 0.7 at $r$=0.98$R$.
Therefore, radiation fields computed from equation (15) 
result in heating rates representative of  sightlines
selected to have arbitrary impact parameters.
Third, 
we computed {\jnu}($0,Z$) at distance $Z$  above the midplane of the
uniform disk and
then averaged the result along sightlines through the disk. This is a more
realistic simulation of the dependence of the heating rate on
mean intensity than estimating {\jnu} at midplane. 
It is encouraging that the 
resulting mean intensities differed by less than  10$\%$ from
the solution in equation (15) for dust-to-gas ratios,
$-$3.0 $<$ log$_{10}${\kapnr}  $<$ 0.0.
We also considered Draine's (1978) solution for {\jnu}($R,Z$) 
in which
the radiation sources are confined to a uniform thin sheet at $Z$ = 0.
The same averaging process led to results in excellent agreement
with equation (15) except when 
log$_{10}${\kapnr} $>$ $-$ 0.5 where the Draine expression
fell significantly below our result. This occurs  because Draine (1978)
excluded sources within a critical radius in the disk location
below the field point in
order to avoid a singularity in his solution.
Fourth, in realistic models of DLAs, {\ps} is not constant throughout
a uniform disk with constant gas density, but rather 
is a function of ${\bf r}$ 
in a system in which gas density also changes with ${\bf r}$.  
We shall examine the implications of this in Paper II.

\subsection{Determining {\ps} in a DLA}

\subsubsection{Technique}

The next step
is to infer {\ps} for each DLA from determinations of {\lclos} and {\kapnr}.  
To do this
we need an additional assumption about the physical state of the gas since
{\lclos} is not a unique function of the other two variables. Indeed, 
for fixed 
[Si/H], {\kapnr}, and {\ps} 
the computed [C II] emission rate per H atom,
{\lcrnr}, varies with  density
(as shown in Figure 3c).
Therefore, to infer {\ps} from  observations
we need to know the density of the gas. 
We address this problem by assuming the gas to be
a two-phase medium with stable CNM gas in pressure equilibrium with
stable  WNM gas. In that case the gas pressure $P$ is restricted to lie
between the local minimum and maximum of the pressure field; i.e., 
$P_{min}$ $<$ $P$ $<$ $P_{max}$ (see Figure 3a). 
For a given {\ps}, {\lcrnr} in 
the CNM is at least ten times larger than {\lcrnr} in the WNM.
Our first model, referred to as CNM, assumes 
that the typical DLA  sightline encounters comparable \\
H I column densities in the CNM and WNM.
Because the empirical quantity {\lclos}, is the density-weighted
average of {\lcr} along the line of sight 
(equation {\ref{eq:lcavlcr}}), {\lclos}
will thus be dominated by contributions
from {\lcr} in the CNM.
As a result,
{\lcrnr} will be insensitive to density, 
as can be seen in 
Figure
3c which shows {\lcrnr} to 
vary by less
than 0.1 dex for 1 $<$ log$_{10}$$n$ $<$ 4 cm$^{-3}$; i.e., in the
density range of the CNM.

This insensitivity to density
is a generic trait of CNM gas,  as shown in Figure 5 where
$P$ and {\lcrnr} are plotted against $n$ for 
a grid of SFRs.
The results for  the Q0458$-$02  
and Q1346$-$03 DLAs are shown in Figures 5a,b and 5c,d respectively.
These were chosen to compare results for a low-$z$
metal-rich DLA and a high-$z$ metal poor DLA.
As {\ps} increases, 
$P_{min}$ and $P_{max}$ increase in magnitude and 
shift to higher densities (see W95).
We shall assume
that the pressure of the two-phase
medium 
equals 
the geometric mean of $P_{min}$ and
$P_{max}$; i.e., $P_{eq}$=$(P_{min}P_{max})^{1/2}$.
In principle, $P_{eq}$ could assume any value between
$P_{min}$ and $P_{max}$. We were guided
by Zel'dovich \& Pikel'ner (1969) who
used stability arguments to derive unique solutions for
$P_{eq}$. Recent numerical simulations tend to support these
conclusions  and show that $P_{eq}$ is closer to $P_{min}$ than $P_{max}$
(Kritsuk \& Norman 2002) in approximate agreement with our
criterion. But given all the uncertainties, we shall not pursue
the stability approach here. 
As discussed in $\S$ 3
\begin{figure*}[ht]
\centering
\scalebox{0.5}[0.6]{\rotatebox{-90}{\includegraphics{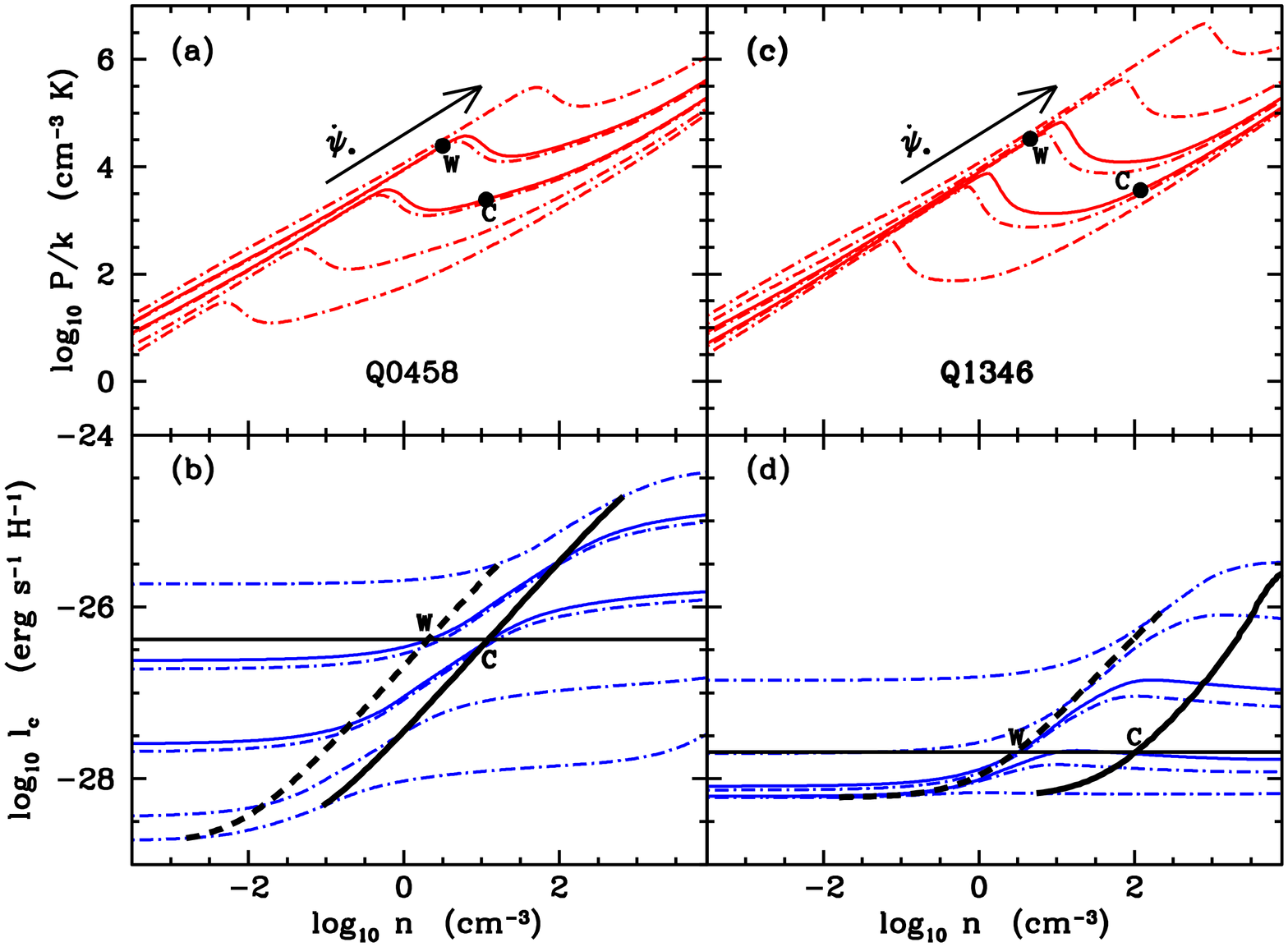}}}
\caption[]{Two-phase diagrams plotting $P/k$ and {\lcrnr} versus density.
Results shown for DLAs toward Q0458$-$02 (Figure 5a,b)
and Q1346$-$03 (Figure 5c,d).
The dot-dash curves in  Figures 5a,c depict
$P(n)$ 
equilibrium solutions for the SFR grid
log$_{10}${\ps} = $-$4,$-$3,...,0 {\smpykpc}. 
Dot-dashed curves in Figures 5b,d depict {\lcrnr}($n$)
equilibrium solutions for the same SFR grid. The solid (dashed)
black curves cutting across these solutions correspond to
{\lcrnr}($n_{CNM}$) and {\lcrnr}($n_{WNM}$) for the SFR grid,
where $n_{CNM}$ and $n_{WNM}$ are thermally stable densities
for which the pressure $P$=($P_{min}P_{max})^{1/2}$. The intersections
between these curves and the observed {\lclos} (horizontal line) yield $n_{CNM}$,
$n_{WNM}$, and the corresponding {\ps} for each DLA. The intersections, denoted
by ``C'' and ``W'', are also shown in the $P(n)$ solutions. The two solid blue curves
in Figures 5b and 5d are the unique solutions passing through the
intersections.} 
\label{Pandlc}
\end{figure*}

\noindent the intersection between $P_{eq}$ and the equilibrium
curve $P(n)$ results in two thermally stable roots:
$n$=$n_{CNM}$ in the high-density CNM,  
and $n$=$n_{WNM}$ in the 
low-density WNM (see also Figure 3). The steeply rising
black solid curves in Figures 5b,d 
connect the $n_{CNM}$, {\lcrnr}($n_{CNM}$) pairs 
determined for each {\ps} of the SFR grid,
where {\lcrnr}($n_{CNM}$) represents {\lcrnr} evaluated
at the density, $n_{CNM}$ determined for a given {\ps}.
The intersections between these
curves and the observed {\lclos} (shown as horizontal lines)
determine $n_{CNM}$ and {\ps} for each DLA. These are denoted by
``C'' in Figures 5b,d as are the corresponding locations
in the $P,n$ plane. The 
unique {\lcrnr}($n$) curves passing through the intersection point
``C'' are  shown as solid blue curves,
and correspond to log$_{10}${\ps} = $-$1.90 and
$-$2.75 {\smpykpc} for Q0458$-$02 and Q1346$-$03.
In most cases the
precise location of $n_{CNM}$ is unimportant as the {\lcrnr} versus
$n$ curves are so flat in the CNM. 
Therefore, the SFRs we derive are relatively insensitive to expressions for
$P_{eq}$ (see discussion in Paper II).
As discussed in $\S$ 5.1 the heating rates averaged along the line of sight
equal the heating rates at midplane to an accuracy better than 
10$\%$. It follows from our steady-state assumption that the
same is true for the cooling rates, and it is this which
justifies the approximation, {\lclos}={\lcrnr}.

Figure 5 also 
illustrates why Norman \& Spaans (1995) suggested that
all high-$z$ DLAs comprise neutral gas only in the WNM phase. 
Suppose {\lclos} for the metal-poor DLA toward Q1346$-$03 ([Si/H]=$-$2.332)
were increased to the lower limit placed on {\lclos} for the
more metal-rich DLA toward Q0458$-$02 ([Si/H]=$-$1.185); 
i.e., log$_{10}${\lclos}=$-$26.38 erg s$^{-1}$ H$^{-1}$.
In that case the SFR implied for the Q1346 DLA would be log$_{10}${\ps}
= $-$1.2 {\smpykpc} which is 5 times the SFR implied by the
same {\lclos} for the Q0458 DLA. The pressure inferred for the 
Q1346$-$03 DLA would increase from 3{$\times$}10$^{3}$ K cm$^{-3}$ to
10$^{5}$  K cm$^{-3}$. 
The Norman \& Spaans (1997) models required
high pressures at high redshifts, since they
assumed  
metallicity to decline rapidly with redshift.
They concluded that the 
gravitational fields generated by  low-mass galaxy progenitors
in CDM models could not supply hydrostatic
pressures as high as 10$^{5}$ K cm$^{-3}$ at $z$ $\sim$ 3.
To solve this problem
these authors concluded  
that {\em all} the neutral gas in high-redshift
DLAs must be low pressure matter in
which $P_{eq}$ $<$ $P_{min}$; i.e., the gas is
a pure WNM. Though
recent studies show that the metallicities of high-$z$ DLAs
are {\em not} as low as assumed by Norman \& Spaans
(see PW02),
there are other, independent arguments for DLAs comprised
of pure WNM gas. Specifically, the failure to detect 21 cm
absorption
in high-$z$ DLAs with large H I column densities led
Kanekar \& Chengalur (2001) to invoke high spin temperatures
as the explanation.
Moreover, Liszt (2002) claimed the 
large CII/CI ratio detected in many DLAs as evidence for WNM gas.

For these reasons we consider the 
alternative hypothesis  that {\ciis} absorption in DLAs originates
in the WNM. More specifically we suppose that all low-ion transitions
in DLAs, such as {\lya}1215, Si II 1527, Fe II 1608, etc., arise in 
low density gas with $T$ $\approx$ 8000 K. 
According to the Norman \& Spaans (1997) hypothesis 
this occurs because
$P$ $<$ $P_{min}$; i.e., CNM gas does not exist in high-$z$ DLAs.
However, the detection of 21 cm absorption with spin temperatures,
$T_{s}$ $<$ 600 K in Q0458$-$02 (Wolfe {\etal} 1985) and $T_{s}$
$<$ 1200 K in Q1331$+$17 (Wolfe \& Davis 1979; Chengalur \& Kanekar 2001)
rules out a pure WNM  and is consistent with the presence of CNM in some cases.
As a result we retain the two-phase hypothesis, but assume the
CNM covering factor is so low that many
sightlines  miss the CNM phase and encounter
only the WNM phase. 
Because we assume $P=P_{eq}$,
the density of the WNM would be given by the thermally stable root
$n_{WNM}$ discussed above. In this case, the steeply rising dashed curves
in Figure 5b,d
connect the $n_{WNM}$,{\lcrnr}($n_{WNM}$) pairs, and their intersections
with the observed {\lclos} are denoted by ``W''.
Proceeding by analogy with the CNM model we find that log$_{10}${\ps}
=$-$0.90 and $-$1.75 {\smpykpc} for the Q0458$-$02 and Q1346$-$03
DLAs respectively.
Because the {\lcrnr} versus $n$ curves are flat at densities below
our solution for $n_{WNM}$, the precise location
of $n_{WNM}$ is not essential in this case either.
Note, the shape of the {\lcrnr} versus $n$
curves are flat in the WNM because at low densities, 
the FUV radiation field
dominates the population of the $^{2}P_{3/2}$ 
and $^{2}P_{1/2}$ fine-structure states
in CII through optical pumping. The
increase of {\lcrnr} with {\ps} is due to the increase 
in the pumping rate caused by the increase in FUV radiation.
By contrast, the high gas densities in the CNM cause {\lcrnr}  
to equal the heating rate. So, in that case
{\lcrnr} increases with {\ps} because the  heating rate
rises  with increasing 
{\ps}. The effects of CMB radiative excitations should be recognizable at
low {\ps}, where pumping is negligible, and at high $z$ 
where the CMB intensity is high. They are evident
in the ($n$, {\lclos}) plane for the Q1346$-$03 DLA
as the sharp flat cutoff
at log$_{10}${\lclos} $<$ $-$ 28 erg s$^{-1}$ H$^{-1}$.
A more detailed discussion of these effects is given in 
Paper II.

Figures 5b,d show that the WNM solutions (points ``W'') require
lower densities and higher SFRs than the CNM solutions (points ``C'').
This generic property of WNM versus CNM solutions is a direct
result of the lower fraction of total cooling due to {\ciis}
in the WNM models. 
Therefore, higher {\ps} are required for {\lcrnr}($n$) curves to
intersect the observed {\lclos} emission rate at the
lower densities characterizing the WNM.
In fact, the {\ps} derived for
the WNM model are conservative lower limits.
Had we used the  
$P$ $<$ $P_{min}$ criterion,  the 
``W'' intersections would occur at  
lower densities than indicated in Figure 5.
But, there is even a more fundamental reason
why  higher values of
{\ps} are required for the WNM solutions; i.e.,
the absence of optical pumping.  

Sarazin {\etal} (1979) showed that the rate of optical pumping 
approaches zero,
if the gas is optically thick in 
UV transitions depopulating the ground-term fine-structure
states. Optical pumping occurs when an radiative upward transition $l$$\rightarrow$$k$
is followed by a downward radiative transition $k$$\rightarrow$$u$ to a different
state, $u$. The net effect is to pump state $u$. At sufficiently large optical
depths pumping
ceases because the lines approach detailed balance in which the pumping rates
$P_{lk}$=$P_{kl}$ and $P_{uk}$=$P_{ku}$. In that case the line
intensities are given by the solution to the two level atom,
and the level populations of the $l$ and $u$ states are determined
only by the population rates among the ground terms $G_{lu}$ and
$G_{ul}$. In the case of C II, the ground term is a fine-structure
doublet and only the resonance transitions $l$$\rightarrow$$k$
need be optically thick for pumping to vanish.   
More specifically, Sarazin {\etal} (1979)
show that pumping ceases when the quantity, $\xi$=
2{$\times$}[$\ln$(${\tau_{lk}}$)/$\pi$)]$^{1/2}$[1+($P_{lk}/G_{lu}$)]
is much less than 1, where $\tau_{lk}$ and $P_{lk}$ specify
the optical depth and transition rates from the $^{2}P_{1/2}$ state
to higher lying states, and $G_{lu}$ specifies collisional and
radiative transition rates for the $^{2}P_{1/2}$$\rightarrow$
$^{2}P_{3/2}$ transitions. The only $l$$\rightarrow$$k$ transitions
relevant for pumping 
are C II 1036.3 and C II 1334.5, since the remaining resonance
transitions, C II 858.1, C II 903.6, and C II 904.0,  have
wavelengths less than 912 {\AA} where {\jnu} will be negligible
compared to the FUV intensities that we derive. In $\S$ 5.2.2
we compute $\xi$ to check
the WNM solutions for self-consistency.

\begin{figure*}[ht]
\centering
\scalebox{0.5}[0.5]{\rotatebox{-90}{\includegraphics{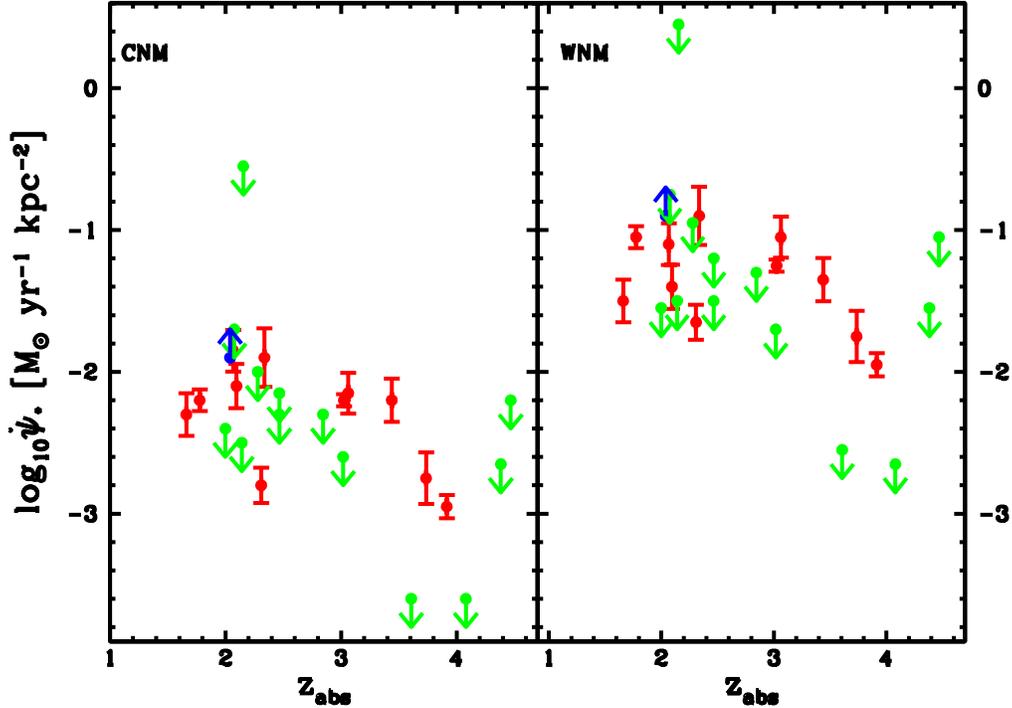}}}
\caption[]{DLA SFR per unit area versus absorption redshift
deduced for dust with ``Gal'' composition and minimal 
depletion. (a) shows results for the CNM model,
and (b) for the WNM model}
\label{psivsz}
\end{figure*}

\subsubsection{Results}

Using the above technique, we determined {\ps} for a subset of 
the 33 DLAs in Table 1. The
DLAs toward Q0951$-$04, Q1425$+$60,
and Q1443$+$27 were omitted since in these objects
only limits  were placed
on [Si/H] or [Fe/H].  
In principle, we could have 
evaluated {\kapnr}, [Si/H], and [Fe/H]  at the limits we have determined.
We rejected this
procedure owing to the sensitivity of the inferred {\ps} to {\kapnr}
and to the sensitivity of the computed {\lcrnr} versus 
$n$ curves to metallicity.
The DLA toward Q0952$-$01 was excluded
because no measurement of [Si/H] or of a relevant proxy  exists. 
We threw out the
DLAs toward Q1759$+$75 and Q2343$+$12, because of evidence that
C$^{+}$/C $<$ 1 and H$^{0}$/H $<$ 1; i.e., the gas
causing damped {\lya} absorption is significantly ionized
(Prochaska {\etal} 2002a; Dessauges-Zavadsky {\etal} 2003). This is in
contrast to most DLAs where H$^{0}$/H{$\approx$}1 (Vladilo {\etal} 2001).
As a result, the maximal
depletion subsample comprises the remaining 27 DLAs. For the
minimal depletion sample we also excluded the DLAs toward Q0201$+$11 and
Q2344$+$12 because the observed [Fe/Si] exceeds the assumed ``nucleosynthetic
ceiling'' value of [Fe/Si]$_{int}$=
$-$0.2. Consequently, the minimal depletion subsample comprises the remaining
25 DLAs. The DLAs excluded from the subsamples are noted in Table 1.

\begin{table*}[ht] {\small} 
\begin{center}
\caption{{\sc Average SFR per Unit Area}}
\begin{tabular}{lcccc}
\tableline
\tableline
\cline{2-5}
&\multicolumn{4}{c}{$<$$\dot{\psi_{*}}$$>$}$^{a}$ \\
\cline{2-5}
&\multicolumn{4}{c}{{\smpykpc}} \\
\cline{2-5}
&\multicolumn{2}{c}{CNM}&\multicolumn{2}{c}{WNM}  \\
\cline{2-3} \cline{4-5}
Dust Model &$z$=2.15$^{b}$&$z$=3.70$^{c}$&$z$=2.15&$z$=3.70   \\
\tableline
``Gal''$^{d}$, max$^{e}$& (3.29$\pm$0.71)10$^{-3}$&(2.23$\pm$0.40)10$^{-3}$&(3.39$\pm$0.66)10$^{-2}$&(2.50$\pm$0.45)10$^{-2}$  \\
``Gal'', min$^{f}$& (7.93$\pm$1.60)10$^{-3}$&(4.52$\pm$1.12)10$^{-3}$&(6.47$\pm$1.21)10$^{-2}$&(4.38$\pm$0.86)10$^{-2}$  \\
``SMC''$^{g}$, max& (9.29$\pm$1.97)10$^{-3}$&(4.45$\pm$1.02)10$^{-3}$&(4.94$\pm$0.99)10$^{-2}$&(3.34$\pm$0.57)10$^{-2}$  \\
``SMC'', min& (1.32$\pm$0.27)10$^{-2}$&(6.39$\pm$1.79)10$^{-3}$&(7.96$\pm$1.61)10$^{-2}$&(4.84$\pm$0.99)10$^{-2}$  \\
\tableline
\end{tabular}
\end{center}
\tablenotetext{a}{Entries are SFRs per Unit Area}
\tablenotetext{b}{Mean redshift of low--$z$ bin}
\tablenotetext{c}{Mean redshift of high-$z$ bin}
\tablenotetext{d}{Carbonaceous ``Gal'' dust}
\tablenotetext{e}{Maximal model where ${\kappa} = 10^{[{\rm Si/H}]_{int}}{\Bigl (}10^{{\rm [Fe/Si]}_{int}}-10^{[\rm Fe/Si]_{gas}}{\Bigr )}$, [Fe/Si]$_{int}$=0}
\tablenotetext{f}{Minimal model where ${\kappa} = 10^{[{\rm Si/H}]_{int}}{\Bigl (}10^{{\rm [Fe/Si]}_{int}}-10^{[\rm Fe/Si]_{gas}}{\Bigr )}$, [Fe/Si]$_{int}$=$-$0.2}
\tablenotetext{g}{Silicate ``SMC'' dust}
\end{table*}

We 
determined {\ps} for each DLA in these subsamples. Figure 6 shows the resulting
{\ps} plotted against redshift 
for both the  CNM and WNM models in the case 
of minimal
dust-to gas ratios and ``Gal'' dust composition. For either
the CNM or WNM models 
the results are qualitatively similar for all four
combinations of maximal or minimal dust-to-gas
ratios and ``Gal'' or ``SMC'' composition.
Due to the lower fraction of cooling carried
by [CII] 158 {\micron} emission in the WNM,
the  star formation rates are at least 10 times
higher for the WNM model than for the CNM model.
Though the positive detections exhibit an apparent decrease
in {\ps} with redshift in the interval $z$=[3,4], no statistically
significant evidence for redshift evolution exists.
A Kendall tau test using only positive detections
shows $\tau$=$-$0.26, while
the probability for the null hypothesis of no correlation, $p_{Kendall}$
=0.26.
Since there is no empirical evidence that the DLAs with limits are
physically distinct from those with measured {\ps}, we shall assume
that all systems are drawn from the same underlying population.
The lowest upper limits on {\ps} at $z$=3.608 
and $z$=4.080 are
possible exceptions to this rule, as these points are outliers
with {\ps} systematically below the range of the main population.
A self-consistent interpretation we shall explore
is that the underlying population of  DLAs consists of 
two-phase media with log$_{10}${\ps}
ranging 
between $-$3 and $-$2 {\smpykpc}. In most cases
the sightlines pass through CNM and WNM gas. 
However, the sightlines through
the two ``outliers'' (the DLAs toward Q1108$-$07 and Q2237$-$06)
pass only through WNM gas in which case the SFRs fall within
the range of the underlying population (see Figure 6b).
If future observations reduce the upper limits on $N$({\ciis}) 
significantly, we would reject this hypothesis and attribute
virtually all of {\lcrnr} in these two systems to excitation by the CMB 
(see Paper II).

Finally, we evaluated the pumping parameter, $\xi$, for the 25 DLAs in the
minimal-depletion ``Gal'' sample. For \\
C II 1036.3 and C II 1334.5 we found that the WNM solutions
for {\ps} resulted  in $\xi$ $<$ 0.4 for the 11 cases of positive {\ciis} detection and
that $\xi$ $<$ 0.2 for 8 of these. We also found $\xi$ $>$ 0.025  for the 
DLA with a lower limit
on {\ciis}. Therefore, {\lcrnr}($n$) is 
likely to be lower in the WNM, and approach the cooling
rate, $n{\Lambda_{\rm CII}}$ (shown as the dotted blue curve
in Figure 3c).
When we recomputed {\ps} in the absence of optical pumping, we found that
{\ps} for the WNM increased between 0.2 and 0.3 dex above the values
shown in Figure 6b. While a more realistic treatment of
optical pumping (see
Flannery {\etal} 1979, 1980), is necessary  to compute 
accurate values of {\ps} in the WNM, it is obvious that the
true values for {\ps} lie somewhere between the  
values shown in Figure 6b and those computed in the limit
of zero pumping rates. In what follows we 
adopt the ``optically
thin'' solutions
with pumping shown in Figure 6b. We shall re-examine
the effects of negligible pumping
in Paper II where we compute the bolometric background radiation
generated by the WNM solutions.

\subsection{The Average SFR per unit area, $<${\ps}(z)$>$}

We now determine the {\em average} star formation rate per area
{\psav} from our sample distribution of {\ps}. 
This is an important statistic since
as we show in Paper II the star formation rate per unit
comoving volume is proportional to {\psav}.
Our goal 
is to 
determine {\psav} in as many redshift bins
as possible.  This is because
we wish 
to determine the star formation {\em history} of DLAs,
and because we wish to compare our results with 
comoving star formation rates 
inferred from flux-limited samples
of galaxies for multiple redshift bins (e.g. Steidel
{\etal} 1999). 
We find that 
dividing the data into more than two redshift bins results in
statistical errors in {\psav} that exceed the  systematic errors. As a result
we split 
the data set at the median redshift, $z$ = 2.7, and determine {\psav}
in two redshift bins with redshift intervals $z_{1}$=[1.6,2.7] and
$z_{2}$=[2.7,4.6].

While positive measurements of {\ps} were inferred
for many DLAs with detected {\lclos},
upper limits on {\ps} were set for an even larger number of 
DLAs with upper limits on {\lclos}, and lower limits were
set on {\ps} for a single DLA
with a lower limit on {\lclos}.
The numbers for the minimal and maximal depletion subsamples
are 11 and 12 positive measurements, 13 and 14 upper limits, and
a single lower limit, respectively.
The presence of large numbers of limits among the data sets presents
a challenge in estimating {\psav}. The arithmetic mean is particularly
sensitive to possible large values in the system for which  only a lower
limit could be measured. As discussed in $\S$ 5.2 
we do not have empirical evidence that the
systems with measured limits are physically distinct from those where the
SFR has been detected. We therefore proceed from the assumption
that the points with limits have been drawn from the same
underlying distribution of SFRs as the detections. We use
the detections to model this distribution empirically, as there is 
no consensus physical model for SFRs in
DLAs at this redshift. We then treat the upper and lower limits
as being drawn randomly from this empirical distribution truncated at the
observed limit value. The mean value of the remaining probability distribution
function is assigned to the data point, and the arithmetic mean of the full
data set including upper and lower limits is then calculated.
We have performed this calculation in two ways, using (1) the observed
distribution of the detections and (2) a Gaussian in log$_{10}$({\ps})   
fit to this observed distribution as our PDF which is then truncated by
the observed limit. The second approach is designed to include a reasonable
probability of high-valued outliers to which the mean is particularly sensitive,
although the mean derived by this method does not differ strongly from the
first method, and indeed neither mean differs strongly from the simple
arithmetic mean of the detections alone. The uncertainty in
the mean of our sample is then
calculated using bootstrap resampling (cf. Efron \&
Tibshirani 1986). The
bootstrap errors are larger than a nominal propagation of the errors on
the individual detections because of significant scatter in
the {\ps}.

We computed {\psav} with the two truncation approaches as well as assuming it
to be the arithmetic mean of the positive detections. The results
of the three techniques
agree within the 1{$\sigma$} errors indicating that {\psav} is a robust
statistic. As a result we henceforth assume {\psav} to be given by
the mean of the positive detections.  
Table 3 shows {\psav} and the 1-$\sigma$ errors
adopted for the CNM and WNM
models in the two redshift bins, for the assumptions 
of maximal and minimal depletion, 
and ``Gal'' and ``SMC'' dust composition. The 
errors are the quadratic sums of the bootstrap
errors and errors of individual detections. 
Obviously the final uncertainty in {\psav} is dominated
by the systematic variation of the mean amongst the models. 
Averaging over the entries in Table 3 we find that
log$_{10}${\psav}=$-$2.19$^{+0.19}_{-0.26}$ {\smpykpc}
for the CNM model, and
log$_{10}${\psav}=$-$1.32$^{+0.13}_{-0.21}$ {\smpykpc}
for the WNM model. 
In Paper II we will reduce
this systematic uncertainty by considering the 
consequences of cosmologically distributed DLAs with the derived SFRs.

\section{CONSTRAINTS ON DUST COMPOSITION IN DLAs}

Having determined the average SFRs per unit area
we now examine the assumptings underlying
the composition  of dust to see whether they
are self consistent.

The heating rate, $\Gamma_d$, is proportional to $\epsilon G_0 n_{grain}/n_H$,
 the product of the photoelectric heating efficiency of 
the dust grains, the FUV radiation intensity, and the 
abundance of the dust grains that dominate the heating.
In $\S$ 4.1 and the Appendix we used 
the dust-to-gas ratio $\kappa$ to determine the abundance of grains 
that dominate the heating for both the Gal and SMC models.  
Our analysis implicitly assumes that the number of depleted C or 
Si atoms per depleted atom of Fe is the same in DLAs as 
in the Milky Way (see equation {\ref{eq:CtoFedepleted}}).
We will now test this assumption for consistency 
with the lack of evidence of Si depletion in DLAs and 
with our assumption that Si and C are undepleted when  
calculating $\kappa$ and the gaseous carbon abundance.

The depletion of Fe was determined by assuming that Si was undepleted.
In that case the depleted ratio [Fe/Si]$_{deplete}$ \\
=[Fe/Si]$_{gas}$
$-$[Fe/Si]$_{int}$ where the intrinsic ratio [Fe/Si]$_{int}$ = 0
in the case of maximal depletion and [Fe/Si]$_{int}$ = $-$0.2
in the case of minimal depletion. 
Since [Fe/Si]$_{gas}$ typically equals $-$0.3, we have
[Fe/Si]$_{deplete}$ = $-$0.3 
for maximal depletion, and [Fe/Si]$_{deplete}$ = $-$0.1 for
minimal depletion. 
Comparison with the abundance of Zn and S implies that Si 
cannot typically be depleted by more than 0.1 dex (PW02). 
Over the range of 
[Fe/Si] observed, this implies that measuring the relative abundance 
of Fe versus a nonrefractory element with the same nucleosynthetic 
history as Si such as S would generate values of  [Fe/S] about 
0.1 dex lower than those observed for [Fe/Si]. This implies that
our technique of using Si as an undepleted element results in 
an underestimate of the dust-to-gas ratio. This is because
the ratio of {\kapnr} based on Si to {\kapnr} based on
S is given by 

\begin{equation}
{{\kappa}_{{\rm Si}} \over {\kappa}_{{\rm S}}}=10^{[\rm Si/S]}{\Biggl (}{{1-10^{[{\rm Fe/Si}]_{gas}-[{\rm Fe/Si}]_{int}}} \over {1-10^{[{\rm Fe/S}]_{gas}-[{\rm Fe/Si}]_{int}}}}{\Biggr )}
\cmma
\label{eq:kapSitokapS}
\end{equation}

\noindent where we assumed [Fe/Si]$_{int}$ = [Fe/S]$_{int}$ since
Si and S are both $\alpha$ enhanced elements. 
Assuming [Si/S]=$-$0.1 we find
we have underestimated the 
dust-to-gas ratios by factors between 1.5 (maximal depletion) and 2.2
(minimal depletion), 
offering the possibility of 
reducing the SFRs by a factor of two in both the Gal and SMC 
models (since our estimate of $\kappa$, i.e., ${\kappa_{\rm Si}}$,
is used to predict the 
dust grain abundance in both models). 

For the purposes of determining the gaseous carbon 
abundance in DLAs, we assumed that carbon 
was undepleted and set [C/H]$_{gas}$=[Fe/H]$_{int}$.  
In the Gal model, however, we are relying on carbonaceous grains to dominate 
the heating, so carbon must be {\it somewhat} depleted.  
Empirical determinations of the C abundance in DLAs are not available 
since all detectable CII resonance lines are saturated
(except in one system).  In our Galaxy, it appears that at least half of 
the C atoms are depleted at all densities 
(Meyer 1999). 
If this is also true in DLAs 
it would reduce the gaseous C abundance by half and alter our thermal
balance solutions such that
$<${\ps}$>$ is reduced by $\approx$ 0.2 dex; this is
a mild change in our results that leads to no qualitative
differences in our conclusions.  It seems more likely that the overall 
carbon depletion level is lower in DLAs due to the reduced metallicity.
In particular, the SMC model has a reduction in the abundance of 
carbonaceous grains and therefore it is likely that the vast majority 
of carbon is gaseous.

Without much knowledge of the composition of dust outside our own 
Galaxy, it is difficult to estimate the systematic uncertainty 
introduced by our assumption that the number of depleted Si and C 
atoms per depleted Fe atom is the same in DLAs as in the Milky Way.  
If there really is a base level of C depletion independent of density 
and metallicity, this implies that using $\kappa$ underestimates the 
true number of depleted carbon atoms.  However, this may still give a reasonable 
estimate of the small carbonaceous grains which dominate the heating 
but do not appear to be part of the base depletion in our Galaxy 
(Sauvage \& Vigroux 1991). 
The uncertainty in the abundance of small carbonaceous grains is 
bracketed by the range of models for the size distribution of 
dust grains in Weingartner \& Draine (2001b).  
The fraction of depleted C atoms in small carbonaceous grains could 
be a factor of two lower than implied by the extrapolated MRN size
distribution
used by our adopted Bakes \& Tielens (1994) model, which 
would reduce 
the photoelectric heating efficiency by a factor of two and thereby 
increase our inferred star formation rates by a factor of two.  
Reducing the overall number of depleted C atoms could make 
the SFRs even higher, but 
reducing the small carbonaceous grain population by more than 
a factor of three makes silicate grains dominate the heating, 
in which case 
the Gal model becomes the SMC model. 
Alternatively, the fraction of depleted C atoms contained in 
small grains could be increased by up to a factor of four.  
If this is the case, or if indexing $\kappa$ to Fe has 
underestimated the carbon depletion, small carbonaceous grains could 
be more abundant than our assumptions imply, leading to higher 
heating efficiency, reduced star formation rates, and a stronger 
2175\AA~ bump.  Observational limits on the strength of 
the 2175\AA~ bump in DLAs make it difficult for small carbonaceous 
grains to be more than a factor of a few  
more abundant than we have 
supposed (Pei {\etal} 1991).    
Tighter observational limits on (and possibly 
detection of) the 2175\AA~ bump in DLAs are of the utmost 
importance in reducing the systematic uncertainties in the nature 
of dust at these redshifts.  

In the SMC model, we suppose the complete absence of small carbonaceous 
grains as inferred from the lack of the 2175\AA~ bump.  
This allows one to lower the depletion level of Si 
considerably and still have silicate grains dominate the heating. 
Since Fe is almost completely depleted in 
diffuse regions of the Milky Way, we are likely to overestimate 
the ratio of $n^{Si}_{depleted}/n^{Fe}_{depleted}$ in a lower 
metallicity region, since our prescription for computing
$n^{Fe}_{depleted}$ in the Appendix would be an underestimate. 
This appears to be the case for the SMC, 
where absorption by clouds along the lines-of-sight 
to Sk 108 and Sk 155 show [Si/Zn]=0, implying a lack of silicon depletion, but 
[Si/Fe]=0.5, indicating that iron is significantly depleted 
(Welty {\etal} 1997; 2001).  
In the diffuse regions of the Galaxy modeled by Weingartner \& Draine (2001b), 
75\% of the Si atoms and 95\% of the Fe atoms are depleted, but in the 
SMC it appears that no more than 20\% of the Si atoms are depleted even 
though 70\% of the Fe atoms are.  This implies that indexing $\kappa$ 
to the Fe depletion overestimates the number of depleted 
Si atoms by at least a factor of three if DLAs are like the SMC.  This 
uncertainty is somewhat balanced by the uncertainty in the 
fraction of Si atoms contained in small ($< 15$~\AA~) 
grains.  In the models of Weingartner \& Draine (2001b), this varies from 
the fraction we have assumed up to a factor of three higher.  
Therefore, 
the abundance of small silicate grains dominating the heating is
unlikely to be
more than a factor of three higher than we have assumed;  if it were, 
the DLA star formation rates for the SMC model would be less than
or equal to 
our current Gal model results.
It is possible, on the other hand, to decrease the 
amount of Si depletion 
by an arbitrary amount, which would 
increase the inferred DLA star formation rates.
For the WNM solution, this would worsen the conflict 
with observational limits on the integrated background 
light discussed in Paper II.  For both WNM and 
CNM solutions, it exacerbates the general
problem of overproduction of metals in DLAs discussed in
Paper II.   
For the CNM solution, doubling
the SMC model star formation rates for DLAs at $z>1.6$ 
begins to exceed the
observational limit on the background
light as discussed in Paper II. 
Hence the integrated background light 
constrains the abundance of small silicate grains in DLAs to 
be at least half of that of the Milky Way.

\section{SUMMARY}

The conventional view of DLAs is they are high-$z$ neutral gas layers with low
metallicities, low dust content, and quiescent velocity fields. 
In this paper we have developed a new technique providing a more
complete picture. Using the
{\ciis} absorption method we find that rather than being 
passive objects transmitting light from background QSOs,
DLAs are the sites of active star formation, and that neutral gas in
DLAs is likely to be a two-phase medium. 
At this stage of our analysis it is unclear whether
{\ciis} absorption arises in the CNM or WNM phase.
Our results are as follows.

(1) Our technique assumes that massive stars forming out
of neutral gas in DLAs emit FUV radiation that heats the gas
by ejecting photoelectrons from dust grains known to
be present in the gas. We can infer the heating rate
since in steady state conditions it
equals the cooling rate that is directly measurable.
This is because cooling is 
dominated by   [C II] 158 {\micron}
emission if the gas is a cold neutral medium (CNM), and
[C II] 158 emission per H atom can
be obtained by measuring {\ciis} 1335.7 absorption
arising from the $^{2}$P$_{3/2}$ excited fine-structure state
in the ground term in C$^{+}$.
The heating rate equals the product of the dust-to-gas 
ratio, the mean intensity of FUV emission, $G_{0}$, and the
grain photoelectric heating efficiency, $\epsilon$.
We can  measure
$G_{0}$ since the cooling  rate is inferred directly from the
column density $N$({\ciis}),
the dust-to-gas ratio can be computed from element abundance patterns, and 
$\epsilon$ is well determined provided the gas is CNM. By measuring
$G_{0}$, we measure
the SFR per unit area, {\ps}, since {\ps} $\propto$ $G_{0}$ in a plane parallel layer.

(2) We have measurements of {\ciis} 1335.7 absorption 
in 33 DLAs; 16 of these are positive detections, 15 are upper
limits, and 2 are lower limits. We use these data to infer the spontaneous
energy emission rate per H atom, {\lclos},
from the ratio $N$({\ciis})/{\NH}. We find that {\lclos} in
DLAs is typically about 1/30 times {\lclos} measured for the
ISM of the Galaxy. Because {\lclos} equals the cooling rate
in the CNM, in our model {\lclos} 
equals the heating rate, which is proportional to  {\kapnr}{$\epsilon$}$G_{0}$,
where {\kapnr} is the dust-to-gas ratio in DLAs relative to 
the ISM.  Since {\kapnr} is also about 1/30, the implication
is that $G_{0}$ in
DLAs is similar to  that in the ISM. In other words, {\ps} in DLAs
is similar to
the local {\ps} in the ISM, provided {\ciis}
absorption arises in a CNM. 

(3) We compute thermal equilibria of gas subjected to grain photoelectric
heating and standard cooling processes. 
Since the dust content of DLAs is
not well determined, we consider
a ``Gal'' model in which the grains are mainly carbonaceous,
and heating is dominated by small ($<$ 15 {\AA}) grains
and PAHs, and an SMC model in which heating is dominated
by small silicate grains. We compute {\kapnr} from the
observed depletion of 
Fe in each DLA and assume that the number of
C or Si atoms depleted onto grains per depleted Fe atom is the
same in DLAs as in the Galaxy. We also consider models with
minimal and maximal depletion to account for the uncertainties
in the Fe depletion levels. We include
heating and ionization due to cosmic rays and soft X-rays. 
When computing cooling rates we account for excitation of
the C$^{+}$  fine-structure levels by
CMB radiation. We also account for excitation
due to optical pumping by FUV radiation, but 
find that optical pumping may not be significant 
owing to the high opacity in the C II resonance lines. 
Equating heating and cooling rates, we find the resulting equilibrium
curves of pressure versus density to exhibit a maximum pressure,
$P_{max}$, and minimum pressure, $P_{min}$. A two-phase medium
in which a dilute warm neutral medium (WNM) is in pressure equilibrium
with a dense CNM is possible if the equilibrium pressure, $P_{eq}$,
satisfies the constraint $P_{min}$ $<$ $P_{eq}$ $<$ $P_{max}$.
Therefore, {\ciis} absorption can occur in the WNM or the CNM.
Because {\lclos} equals the cooling rate in the CNM, but is
a small fraction of the cooling rate in the WNM, the SFRs
implied from a measured {\lclos} are much higher in the 
WNM than the CNM.

(4) We calculate {\ps} for selected subsets of our DLA sample
corresponding to ``Gal'' or ``SMC'' dust, and to minimal or
maximal depletion. We first
solve the transfer equation for sources of FUV ($\approx$ 1500 {\AA})
radiation (OB stars)
uniformly distributed throughout a plane parallel disk. Using standard
reddening curves for ``Gal'' and ``SMC'' dust we find the disks
to be optically thin parallel to the plane for most of our sample DLAs.
As a result, 
{\em the {\ps} inferred from {\ciis} absorption
are representative of the entire DLA rather than just 
regions along the QSO sightline.}  To infer {\ps}
from measurements of {\lclos} and {\kapnr} we assume the
equilibrium pressure, $P_{eq}$=$(P_{min}P_{max})^{1/2}$. 
As expected,
we find two solutions; one in which {\ciis}
absorption occurs in the WNM,  
and the other in which {\ciis} absorption occurs in 
the CNM. For the CNM solution we find $-$3.0 $<$ log$_{10}${\ps} $<$ $-$
2.0 {\smpykpc}, and $-$2.0 $<$ log$_{10}${\ps} $<$ $-$ 1.0 {\smpykpc}
for the WNM solution. Neither case shows evidence for redshift evolution
in the interval $z$ = [1.6,4.5]. In
Paper II we discriminate between these models by deriving cosmological
constraints such as the bolometric background radiation.

(5) Our assumptions that C and Si are undepleted in determining 
the gaseous C abundance and the dust-to-gas ratio are reasonable and 
do not create serious contradictions with the heating being dominated 
by carbonaceous or silicate grains, because the C and Si depletion 
levels in DLAs appear to be quite low.  
Varying the size distribution of carbonaceous grains and the 
number of depleted C atoms per depleted Fe atom 
could 
make the SFRs reported for the Gal model 
as high as those of the SMC model, but one cannot 
lower the SFRs too far since 
the 2175\AA~ bump has not been observed in high redshift DLAs.  Varying 
the number of depleted Si atoms per depleted Fe atom
could make the SFRs 
reported for the SMC model as high as allowed by the integrated background 
limits, and varying the size distribution of silicate grains 
could make the SFRs a factor of three lower.

\acknowledgements

The authors wish to extend special thanks to those of Hawaiian
ancestry on whose sacred mountain we are privileged to be
guests. Without their generous hospitality, none of the observations
presented here would have been possible.  We wish to thank Chris McKee
for many valuable discussions about multi-phase media. We also thank
Bruce Draine, Eli Dwek, Mike Fall, Rob Kennicutt, Alexei Kritsuk, Jim
Peebles, Blair Savage, Marco Spaans, Alexander Tielens, and Mark
Wolfire for valuable comments.  We are grateful to A. Silva and
S. Viegas for sending us their program, POPRATIO, and to Len Cowie,
Wal Sargent, and Tony Songaila for giving us data prior to
publication.  Finally we thank Simon White for remarks that stimulated
this research.  A.M.W. was partially supported by NSF grant AST
0071257.

\appendix

\section{DUST TO GAS RATIO}

Consider a box containing $N_{Fe}^{Tot}$ iron atoms. If
$N_{{\rm Fe}}^{Gas}$ atoms are in the gas phase and $N_{{\rm Fe}}^{Dust}$ 
atoms are locked up in grains, then
\begin{equation}
N_{{\rm Fe}}^{Dust}=N_{{\rm Fe}}^{tot}-N_{{\rm Fe}}^{gas}
\perd
\end{equation}

\noindent Because we cannot measure $N_{{\rm Fe}}^{tot}$ due
to depletion of Fe atoms onto grains, we shall use Si as a proxy as
Si is essentially undepleted in DLAs (see PW02). Therefore,
we assume $N_{{\rm Fe}}^{Tot}$=$N_{{\rm Si}}({\rm Fe/Si})_{int}$ where $N_{{\rm Si}}$,
the total number of Si atoms also equals the number of gas-phase Si atoms,
and $({\rm Fe/Si})_{int}$ is the intrinsic (i.e., undepleted) ratio
of Si to Fe. Let the  dust-to-gas ratio $k$
{$\equiv$}$N_{{\rm Fe}}^{Dust}$/$N_{{\rm H}}$. As a result 
\begin{equation}
k={{N_{{\rm Si}}({\rm Fe/Si})_{int}} \over N_{{\rm H}}}-{N_{{\rm Fe}}^{Gas} \over N_{{\rm H}}}
\cmma
\end{equation}

\noindent where $N_{{\rm H}}$ is the number of H atoms. Since
{\kapnr}$\equiv$
$k/k_{MW}$, where $k_{MW}$ is the dust-to-gas ratio
of the current Milky Way Galaxy, and assuming that $k_{MW}$=(Fe/H)$_{\odot}$,
since Fe is almost entirely depleted,
we find
\begin{equation}
{\kappa}={{({\rm {Si}/{H}})({\rm Fe/Si})_{int}} \over ({\rm Fe/H})_{\odot}}-{{({\rm Fe/H})^{Gas}} \over {({\rm Fe/H})_{\odot}}}
\perd
\end{equation}

\noindent Factoring Fe/H=(Fe/Si)(Si/H)
 we find
\begin{equation}
{\kappa}={\rm {(Si/H) \over (Si/H)_{\odot}}{{(Fe/Si)_{int}} \over (Fe/Si)_{\odot}}-{{(Si/H)(Fe/Si)^{Gas}} \over {(Si/H)_{\odot}(Fe/Si)_{\odot}}}}
\perd
\end{equation}

As a result
\begin{equation}
{\kappa}={\rm 10^{[Si/H]_{int}}{\biggl [}10^{[Fe/Si]_{int}}-10^{[Fe/Si]_{gas}}{\biggr ]}}
\cmma
\end{equation}

\noindent where we assume [Si/H]=[Si/H]$_{int}$.

\newpage

\end{document}